\documentclass[preprint,3p,sort&compress]{elsarticle}
\usepackage{graphicx}           
\usepackage{amssymb,amsmath}
\usepackage{xspace}
\usepackage{color}
\usepackage{longtable}
\usepackage[shortlabels]{enumitem}
\usepackage{hyperref}
\usepackage{caption}
\usepackage{subcaption}
\usepackage{siunitx}
\usepackage{comment}
\usepackage{epstopdf}
\makeindex             
\newcommand{\nc}{\newcommand}
\nc{\rnc}{\renewcommand}
\nc{\mm}{\boldsymbol}
\nc{\bs}{\boldsymbol}
\rnc{\matrix}[2]{\left[\!\!\begin{array}{#1}
	#2\end{array}\!\!\right]}
\rnc{\vector}[1]{\matrix{c}{#1}}
\nc{\dd}{\mathrm{d}}
\nc{\ii}{\mathrm{i}}
\nc{\ee}{\mathrm{e}}
\nc{\inv}{^{-1}} 
\nc{\herm}{^{\mathrm H}}
\nc{\tra}{^{\mathrm T}}
\nc{\conj}[1]{ \overline{#1} }
\nc{\normal}{\mathrm n}
\nc{\tangential}{\mathrm t}
\nc{\kn}{{k_{\normal}}}
\nc{\kt}{{k_{\tangential}}}
\nc{\ie}{i.\,e.\xspace}
\nc{\eg}{e.\,g.\xspace}
\nc{\vs}{vs.\xspace}
\nc{\cf}{cf.\,\xspace}
\nc{\myquote}[1]{`#1'}
\nc{\etal}{et al.\xspace}
\nc{\etc}{etc.\xspace}
\nc{\ABAQUS}{{\sf{ABAQUS}}\xspace}
\nc{\MATLAB}{{\sf{MATLAB}}\xspace}
\nc{\fabstand}{\,}
\nc{\fp}{\fabstand.}
\nc{\fk}{\fabstand,}
\nc{\tab}[5][tbh]{\begin{table}[#1]\centering\caption{#4\label{tab:#5}}\begin{tabular}{#2}\hline #3 \\ \hline\end{tabular}\end{table}}
\nc{\fig}[4][tbh]{
\begin{figure}[#1]
\centering
\includegraphics[width=#4\textwidth]{#2}
\caption{#3\label{fig:#2}}
\end{figure}}
\nc{\e}[2]{\begin{equation} #1 \label {eq:#2} \end{equation}}
\nc{\est}[1]{\begin{equation*} #1 \end{equation*}}
\nc{\ea}[1]{
\begin{eqnarray}
#1 \end{eqnarray}}
\nc{\east}[1]{
\begin{eqnarray*}
#1 \end{eqnarray*}}
\nc{\fref}[1]{{Fig.~\ref{fig:#1}}}
\nc{\frefo}[1]{{\ref{fig:#1}}}
\nc{\frefs}[1]{{Figs.~\ref{fig:#1}}}
\nc{\tref}[1]{{Tab.~\ref{tab:#1}}}
\nc{\trefo}[1]{{\ref{tab:#1}}}
\nc{\trefs}[1]{{Tab.~\ref{tab:#1}}}
\nc{\eref}[1]{{Eq.~(\ref{eq:#1})}}
\nc{\erefo}[1]{(\ref{eq:#1})}
\nc{\erefs}[1]{{Eqs.~(\ref{eq:#1})}}
\nc{\sref}[1]{{Section~\ref{sec:#1}}}
\nc{\srefo}[1]{\ref{sec:#1}}
\nc{\srefs}[1]{{Sections~\ref{sec:#1}}}
\nc{\ssref}[1]{{Section~\ref{sec:#1}}}
\nc{\ssrefo}[1]{\ref{sec:#1}}
\nc{\ssrefs}[1]{{Section~\ref{sec:#1}}}
\nc{\aref}[1]{{{\ref{asec:#1}}}}
\nc{\arefo}[1]{{\ref{asec:#1}}}
\nc{\arefs}[1]{{{Appendices~\ref{asec:#1}}}}
\nc{\inst}[1]{$^{#1}$}
\nc{\naft}{N}
\nc{\ndof}{N_{\mathrm{DOF}}}
\nc{\Neq}{N_{\mathrm{eq}}}
\nc{\nfact}{N_{\mathrm{fact}}}
\nc{\nsolpt}{N_{\mathrm{pt}}}
\nc{\nord}{P}
\nc{\nnewtavg}{\overline{N}_{\mathrm{newt}}}
\nc{\regeps}{\varepsilon_{\mathrm{reg}}}
\nc{\epstol}{\varepsilon_{\mathrm{tol}}}
\nc{\sign}{\operatorname{sgn}}
\nc{\mrm}{\mathrm}
\nc{\mum}{$µ$\mrm{m}}
\journal{MSSP}

\begin{document}

\begin{frontmatter}

\title{A sub-structuring approach for model reduction of frictionally clamped thin-walled structures}
\author{Patrick Hippold, Johann Gross, Malte Krack}
\address{University of Stuttgart, Stuttgart, GERMANY}

\begin{abstract}
Thin-walled structures clamped by friction joints, such as aircraft skin panels are exposed to bending-stretching coupling and frictional contact.
We propose an original sub-structuring approach, where the system is divided into thin-walled and support regions, so that geometrically nonlinear behavior is relevant only in the former, and nonlinear contact behavior only in the latter.
This permits to derive reduced component models, in principle, with available techniques.
The Hurty-/Craig-Bampton method, combined with an interface reduction relying on an orthogonal polynomial series, is used to construct the reduction basis for each component.
To model geometrically nonlinear behavior, implicit condensation is used, where an original, engineering-oriented proposition is made for the delicate scaling of the static load cases required to estimate the coefficients of the nonlinear terms.
The proposed method is validated and its computational performance is assessed for the example of a plate with frictional clamping, using finite element analysis as reference.
The numerical results shed light into an interesting mutual interaction:
The extent of geometric hardening is limited by the reduced boundary stiffness when more sliding occurs in the clamping.
On the other hand, the frictional dissipation is increased by the tangential loading induced by membrane stretching.
\end{abstract}


\end{frontmatter}

\section{Introduction\label{sec:intro}}
Thin-walled structures have a widespread use in aircraft, space, and wind turbine industries to achieve high strength-to-weight ratios.
Examples are skin panels of wings and fuselage (airplanes, helicopters, space structures), aircraft engine cowlings, and covers or fairings (\eg hand-held power tools, transport systems).
A key feature of these systems is that both frictional contact and geometric nonlinearity are relevant.
More specifically, thin-walled structures (\eg plates, shells, panels, arches, beams) are commonly assembled via mechanical joints using fasteners (\eg pins, rivets, bolts), so that dry frictional and unilateral interactions occur at the contact interfaces.
It is well-known that the frictional interactions in mechanical joints are often the main source of mechanical damping in macroscopic built-up structures \cite{Gaul1997,popp2003a}.
On the other hand, in the case of clamped ends, bending induces membrane stretching/compression, which, in turn, affects the bending stiffness; this bending-stretching coupling is an important example of geometric nonlinearity.
Bending-stretching coupling can have a severe effect on the natural frequencies, and trigger nonlinear modal interactions already for bending deformations in the order of magnitude of the thickness of thin-walled structures \cite{nayf1979,Thomsen.2003,Thomas.2016}.
In other words, the thinner the structures, the more susceptible they are to geometrically nonlinear behavior, while nonlinear frictional contact behavior determines their damping and, hence, whether they survive vibrations.
Based on this, one may expect that both types of nonlinear behavior become increasingly important with the ubiquitous engineering trend towards extreme lightweight design.
\\
Nonlinear kinematic and contact behavior are important in several applications, including rotor-bearing systems (see \eg \cite{Jin.2017,Jin.2019b,Lu.2021}), and blade-casing interaction (see \eg \cite{Xiao.2021,Delhez.2021,Delhez.2023,Mashayekhi.2024}).
The focus of the present work is on frictionally clamped thin-walled structures (or more briefly: thin-walled jointed structures).
For the above reasons, we are convinced that this problem class is of very high engineering relevance.
Experiments have shown the relevance of nonlinear behavior of thin-walled jointed structures, see \eg \cite{Yun.1998,Claeys.2014,Anastasio.2019,Karaagacl.2022,Bhattu.2024}.
Efforts on prediction and validation have been surprisingly scarce for this problem class.
The high engineering relevance of thin-walled jointed structures, and the lack of validated prediction methods was the primary motivation for the Tribomechadynamics Research Challenge \cite{TRCprediction}.
This Challenge has triggered considerable research efforts \cite{TRCprediction,NajeraFlores.2024,Farokhi.2024,Morsy.2025}, and, in particular, revealed the lack of appropriate reduced modeling approaches.
The aim of the present work is to develop a model reduction approach for thin-walled jointed structures.
\\
An interesting mutual interaction of both sources of nonlinear behavior is to be expected:
On the one hand, the extent of bending-stretching coupling depends on the effective axial support stiffness \cite{Ibrahim.2005}, which is highly dependent on the contact interactions and, thus, amplitude-dependent.
On the other hand, one can expect that bending deformation leads to opening contacts, whereas at higher amplitudes, stretching should increase the tangential load on the contact interfaces.
It is important to emphasize the multi-scale character of the problem:
The local relative displacements within a frictional clamping/joint may be in the sub-micrometer range, while the maximum (absolute) vibration level of the jointed structure may be much larger (away from the clamping/joint), \eg, in the order of several millimeters.
Hence, both types of nonlinear behavior can be important at the same time, perhaps at other locations within the structure/system.
\\
In the following, an overview is given on the relevant state of the art of model reduction.
First, sub-structuring and component mode synthesis within linear structural dynamics is described, followed by the nonlinear coupling via contact models.
Then, model reduction of geometrically nonlinear structures is addressed, and some pioneering work on sub-structuring of geometrically nonlinear components is presented.
From the deficiencies of the current state of the art, the purpose of the present work is finally derived and the outline of the remaining article is explained.
Although data-based model reduction enjoys increasing popularity \cite{Lu.2019}, we focus here on equation-driven methods.
The intent behind this is to avoid the need to generate data from the high-fidelity model through costly simulations, and and to make it easier to obtain a large parameter space.

\subsection{Reduced modeling of the dynamics of linear structures: sub-structuring, component mode synthesis}
In the linear case, model reduction is strongly linked to sub-structuring techniques.
Here, the system is divided into smaller components (or substructures), reduced models are derived for each component, and finally the reduced system model is obtained by accounting for the behavior at the interfaces.
An important advantage of sub-structuring is the modular setup:
When only an individual component is exchanged, the reduced model of the modified system is obtained with minimal effort \cite{klerk2008}.
\\
The standard approach for deriving reduced component models from a given finite element model is \emph{component mode synthesis}.
The by far most popular technique is the Hurty-/Craig-Bampton method \cite{crai1968,hurt1965}.
Here, the nodal displacements are approximated as a linear combination of static constraint modes and a set of fixed-interface normal modes.
The \emph{static constraint modes} are the displacement vectors obtained for static unit displacements at the interface.
These modes are needed to account for the behavior at the component interface within the assembly.
The \emph{fixed-interface normal modes} correspond to the eigenvectors obtained under zero displacement at the interface.
Commonly, a \emph{target frequency band} is specified, which spans the relevant spectrum of external and internal forces, and only the small subset of normal modes in that band is retained.
By projecting the dynamic force balance onto this reduction basis, one obtains a reduced-order model.
Several methods analogous to the Hurty-/Craig-Bampton method exist, for instance those relying on free- instead of fixed-interface normal modes \cite{crai1977,macn1971,Rubin.1975,rixe2004}, or the moment matching and Krylov-subspace techniques, see \eg \cite{Holzwarth.2015,Holzwarth.2018}.
The latter have mainly been applied to linear problems, and all of those techniques rely on a specific target frequency band.

\subsection{Modeling of nonlinear coupling between components, reduced contact modeling}
Within the above described approach, nonlinear contact behavior can be easily considered at the interface, in terms of either constitutive or set-valued laws.
A fine contact mesh is often needed to spatially resolve the in general non-homogeneous unilateral and frictional interactions, and to accurately describe their influence on the effective stiffness and damping of an interface.
Contact model reduction methods are under research, which propose an interface reduction, combined with an approximation of the projected contact forces by evaluating the contact stress at a reduced set of contact nodes (referred to as hyper-reduction) \cite{Balaji.2021,Koller.2021,Witteveen.2022,Morsy.2023}.
It seems straight-forward to combine the method proposed in the present work with contact reduction.
But since none of those reduction methods is well-established yet, this is regarded as beyond the scope of this article.
Noteworthy is also the ad-hoc generalization of the Hurty-/Craig-Bampton method proposed in \cite{Joannin.2017,Joannin.2018}, where the set of fixed-interface normal modes is replaced by a single nonlinear mode.
Although the limitations of that method have not been fully explored yet, by its design, it strictly assumes that the response in each component is dominated by a single mode and a single harmonic.
Those are important theoretical restrictions, which we intend to avoid with the method developed in the present work.

\subsection{Reduced modeling of the dynamics of geometrically nonlinear structures\label{sec:NLROM}}
In contrast to contact, geometrically nonlinear behavior is not localized to an interface, but it is distributed within the solid.
This requires dedicated model reduction techniques.
Most of these are incompatible with sub-structuring; \ie, they can only be applied on system-level/to single-component systems.
In the absence of (substructure) interfaces, and associated interface modes, the Hurty-/Craig-Bampton method degenerates to the common modal truncation.
For thin-walled structures, the low-frequency modes are typically bending- and torsion-type modes, whereas membrane-type/stretching modes usually have much higher frequency.
Retaining only the (lowest-frequency) bending modes, and using a conventional Galerkin projection, inhibits membrane motion, and, thus, leads to a very poor approximation of bending-stretching coupling.
To overcome this problem, three main avenues can be pursued according to \cite{Touze.2021}:
\begin{enumerate}[(a)]
    \item use a variable reduction basis,
    \item enrich the (constant) reduction basis, or
    \item use static/implicit condensation.
\end{enumerate}
An important example for a variable reduction basis is an invariant manifold.
The most recent techniques construct this manifold and the corresponding reduced model using a direct parametrization \cite{Breunung.2018,Vizzaccaro.2022,Opreni.2023}.
These techniques are not limited to geometrically nonlinear behavior \cite{Opreni.2023b}.
However, as they rely on Taylor series expansion, they are restricted to smooth behavior, and hence are not expected to be useful for frictional and unilateral contact interactions; \ie, they are limited to ideal boundary conditions only. 
Further, there is no opportunity to extend the basis by interface modes, which is why these techniques cannot be used within a sub-structuring framework; they can only be applied on system level. 
In contrast, when a linear combination of (constant) vectors is used as displacement reduction basis, interface modes can be considered, so that (b) and (c) are generally applicable within the sub-structuring framework.
\\
For the aforementioned remedy (b), in particular, the basis can be enriched by membrane-type modes \cite{Hollkamp.2008}.
While the associated membrane modes can be identified by intuition in the case of a flat beam or plate, using modal derivatives seems better-suited in more complicated settings.
Under the assumption of linear elasticity and a total Lagrangian formulation, geometrically nonlinear terms take the form of cubic-degree polynomials (involving quadratic- and cubic-degree terms) in the nodal coordinates of a solid finite element model, see \eg \cite{Holzapfel.2002}.
Of course, a Galerkin projection with a constant basis yields cubic-degree polynomials also in the generalized coordinates of the reduced model.
The coefficients can be determined directly within an appropriately modified finite element code (referred to as \emph{intrusive} procedure) \cite{Idelsohn.1985, Slaats.1995, Shi.1996, Tiso.2011}.
An important downside of using modal derivatives is that their number grows rapidly with the number of initial component modes (fixed-interface and static constraint modes in the case of the Hurty-/Craig-Bampton method).
This is in contrast to static/implicit condensation, where good results can often be achieved with only a single or a very small number of component modes.
\\
In contrast to the aforementioned remedies (a) and (b), static/implicit condensation does not rely on a Galerkin projection.
The idea of a static condensation is to neglect inertia forces associated with high-frequency modes (\eg membrane modes), so that the associated modal coordinates can be eliminated algebraically \cite{Mignolet.2013}.
Formally, the roots of some cubic-degree multivariate polynomial equations are substituted into other quadratic- and cubic-degree polynomial terms.
This generally does not yield a cubic-degree polynomial in the reduced coordinates.
However, using a cubic-degree polynomial is consistent with a third-order Taylor series expansion around the reference configuration.
Higher-order expansions \cite{Frangi.2019,Nicolaidou.2020,Nicolaidou.2022,Xiao.2023} or piece-wise defined nonlinear terms such as splines are occasionally used with the intent to achieve high accuracy over a larger range of displacements \cite{Touze.2021}.
The aforementioned algebraic elimination is intractable for most interesting problems.
Implicit condensation follows the same idea as static condensation, but estimates the coefficients of the nonlinear terms by regression to a set of static load cases \cite{McEwan.2001,Hollkamp.2008,Gordon.2011}.
The accuracy of this estimation is highly sensitive to the scaling of the load cases \cite{Mignolet.2013,Shen.2021,Xiao.2023}:
If the loads are too small, the estimated coefficients are distorted by numerical noise, and if they are too large, higher-order nonlinear terms are activated, which distorts the estimated lower-order coefficients.
An important limitation of static/implicit condensation is that nonlinear inertia effects within the reduced basis are not modeled \cite{Vizzaccaro.2021}.
Also, the frequency of the eliminated modes must be sufficiently high so that neglecting the associated inertia forces does not lead to a poor approximation \cite{Haller.2017,Shen.2021,Vizzaccaro.2021b}.
This is in contrast to the invariant-manifold based approaches mentioned earlier \cite{Breunung.2018,Vizzaccaro.2022,Opreni.2023}.

\subsection{Sub-structuring of geometrically nonlinear components}
Even though it was stated above that the aforementioned methods (b) and (c) are in principle compatible with sub-structuring, only three publications are known to the authors, which actually propose a sub-structuring approach for geometrically nonlinear components \cite{Kuether.2015b,Kuether.2017,KaramoozMahdiabadi.2019}.
The approach proposed by Kuether \etal \cite{Kuether.2015b,Kuether.2017} is based on the Hurty-/Craig-Bampton method and implicit condensation.
A fundamental challenge of implicit condensation is that the number of static load cases required for the regression, like the number of sought coefficients in the reduced model, grows with the cube of the number of component modes \cite{Mignolet.2013}.
That is why it was proposed in \cite{Kuether.2017} to approximate the static constraint modes using a reduced set of interface modes (\emph{interface reduction}).
Local-level interface modes were used in \cite{Kuether.2017}, which are obtained from an eigenvalue problem involving the reduced mass and stiffness matrix, restricted to the interface partition.
In \cite{KaramoozMahdiabadi.2019}, a free-interface method was used for component mode synthesis, in combination with two types of local-level and system-level interface modes.
It was concluded that all variants perform similar in terms of accuracy and computational effort.
System-level interface modes are obtained from the reduced mass and stiffness matrix of the assembled system.
This somewhat defies the purpose of sub-structuring, as the reduced model of each component has to be recomputed when a single component is exchanged.

\subsection{Purpose and outline of the present work}
The purpose of the present work is to develop a sub-structuring method for model reduction of thin-walled jointed structures, taking into account both geometrically nonlinear and nonlinear contact behavior.
It will be exploited that only the thin-walled region has geometrically nonlinear behavior, while contact interactions occur in the clamping only.
The aim of the reduced-order model is to capture the essential dynamics of the parent full-order model, including asymptotic stability.
The proposed method is presented in \sref{proposition}.
In \sref{numerical}, the method is validated against direct analysis of the (full-order) finite element model, and the computational performance is assessed.
To this end, a plate with frictional clamping is considered.
The numerical analysis of this system demonstrates interesting mutual interactions between the two types of nonlinear behavior.
Besides the amplitude-dependence of the frequency and the damping ratio of the fundamental bending mode, the response to impulsive loading is analyzed, and the benefits of the modular character inherent to sub-structuring is illustrated by replacing a support region.

\section{Proposed method\label{sec:proposition}}
A thin-walled jointed structure is schematically illustrated in \fref{divideJointThinwalledRegion}.
The point of departure for the proposed model reduction method is a finite element model.
The dynamic force balance is expressed as
\ea{
\mm M \ddot{\mm q} + \mm K\mm q + \mm h = \mm f_{\mrm{ext}}(t)\fk \label{eq:FOM}
}
where $\mm q$ is the vector of nodal coordinates,
overdot denotes derivative with respect to time $t$,
$\mm M=\mm M^{\mrm T}>\mm 0$ is the symmetric and positive definite mass matrix, $\mm K=\mm K^{\mrm T}\geq \mm 0$ is the symmetric and positive (semi-) definite stiffness matrix, $\mm h$ is the vector of nonlinear internal forces which model both geometrically nonlinear and contact behavior, and $\mm f_{\mrm{ext}}(t)$ is the vector of external forces with known explicit time dependence.
$\mm q$ counts from the static equilibrium (\emph{reference configuration}) obtained under consideration of the initial geometry, the static forces in the form of \eg bolt tightening and thermal loading.
Consequently, $\mm q$ is interpreted as the response to dynamic external forces $\mm f_{\mrm{ext}}$, compatible with the linear and nonlinear dynamic internal forces (left-hand side of \eref{FOM}); the dynamic forces also count from the corresponding static ones.
\\
The original idea to divide the system into thin-walled and support region is explained in \ssref{divide}.
The reduction of the individual regions, using the Hurty-/Craig-Bampton method with an appropriate interface reduction, is described in \ssref{CMS}.
Within the sub-structuring framework, it will be shown that generic nonlinear contact behavior can be directly modeled at the respective interfaces. 
The geometrically nonlinear behavior is modeled using implicit condensation.
Here, an original approach is proposed in \ssref{IC}, as solution of the delicate problem of scaling the static load patterns, required for the regression of the geometrically nonlinear terms.
The assembly of the reduced component models is described in \ssref{assembly}.
A summarizing overview of the proposed method is given in \ssref{overview}.

\subsection{Division into thin-walled and support regions\label{sec:divide}}
\fig[ht]{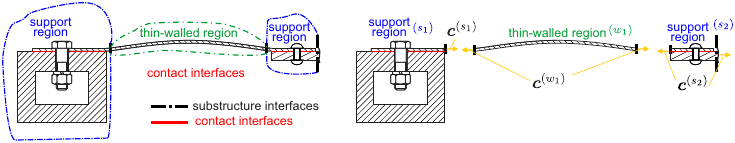}{Proposed sub-structuring approach: (left) division into thin-walled and support regions; (right) definition of coupling forces.}{1.0}
%
The state of the art in sub-structuring is to divide into parts, \eg plate, stiffener, bolt, nut and so on.
The plate would then be exposed to both geometrically nonlinear and nonlinear contact behavior.
In lack of an effective contact reduction method (\cf \sref{intro}), this would lead to a very large number of component modes (associated with the contact interface), making the reduced modeling of geometrically nonlinear behavior computationally infeasible with the available methods.
The \emph{original idea} proposed in the present work is to \emph{divide into thin-walled and support regions}.
This way, the two \emph{types of nonlinear behavior are isolated}:
The support region only exhibits nonlinear contact, the thin-walled region only geometrically nonlinear behavior.
This excludes systems featuring nonlinear contact behavior in the thin-walled region such as sandwich structures with friction between layers.
Since the boundary is at the thin-walled region, the substructure interface is relatively small, which will turn out to be useful for interface reduction.
The resulting components can then, in principle, be reduced with available methods of category (b) and (c) introduced in \ssref{NLROM}.
The proposed sub-structuring approach leads to the situation that the (uncoupled) thin-walled component typically has rigid-body degrees of freedom.
This requires special treatment, as described later.
\\
Before delving into the mathematical formulation of the proposed approach, it is useful to emphasize that the proposed approach enjoys the benefits of the modular setup of all sub-structuring techniques:
The parameters of the individual component models can be varied, or even entire reduced component models can be exchanged, without having to re-derive the reduced models of the other components.
This is an important aspect because setting up the reduced component models is commonly the bottleneck of the analysis.
\\
The above described division splits \eref{FOM} into two sets of equations,
\ea{
{\mm M}^{(s)}\ddot{{\mm q}}^{(s)} + {\mm K}^{(s)}{\mm q}^{(s)} + {\mm f}_{\mrm{con}}^{(s)} + {\mm c}^{(s)} &=& {\mm f}_{\mrm{ext}}^{(s)}(t)\fk\label{eq:supportFOM}\\
{\mm M}^{(w)}\ddot{{\mm q}}^{(w)} + {\mm K}^{(w)}{\mm q}^{(w)} + {\mm f}_{\mrm{geom}}^{(w)} + {\mm c}^{(w)} &=& {\mm f}_{\mrm{ext}}^{(w)}(t)\fp\label{eq:thinwalledFOM}
}
Herein, $\square^{(s)}$ refers to support and $\square^{(w)}$ to thin-walled regions, respectively, for all $s\in\mathcal S$, $w\in\mathcal W$, where $\mathcal S$ and $\mathcal W$ are the sets of support and thin-walled regions, respectively.
${\mm f}_{\mrm{con}}^{(s)}$, ${\mm f}_{\mrm{geom}}^{(w)}$ are dynamic force vectors accounting for contact and geometrically nonlinear behavior.
${\mm c}^{(j)}$ (for all $j\in\mathcal S\bigcup\mathcal W$) account for the coupling of the substructure interfaces.
All other matrices and vectors in \erefs{supportFOM}-\erefo{thinwalledFOM} correspond to the decoupled configuration.
Linear damping forces, \eg in the popular form of modal damping can be easily included into the proposed method.
In the present work, only frictional dissipation is considered.
The equation system \erefo{supportFOM}-\erefo{thinwalledFOM} is to be closed by appropriate models for the contact and the geometrically nonlinear behavior (\ssrefs{CMS}-\ssrefo{IC}), and the coupling conditions (\ssref{assembly}).

\subsection{Derivation of reduced component models\label{sec:CMS}}
To construct the reduced basis of each component, the method proposed by Carassale and Maurici \cite{Carassale.2017} is used as point of departure in the present work.
It augments the well-known Hurty-/Craig-Bampton method with an interface reduction relying on an orthogonal polynomial series.
An important benefit of this method over using local-level interface modes is that it avoids the computation of the complete set of static constraint modes, and the associated reduced stiffness and mass matrices.
This method is applied to reducing nonlinear models for the first time here.
To make this article self-contained, the method is briefly described in the following.
Also, an illustration of the interface mode shapes is useful for understanding the modal convergence analyzed in \sref{numerical}. 
As the reduced basis is computed individually for each component, the specifiers $\square^{(s)}$, $\square^{(w)}$ are omitted in the following; they are used again upon assembly (\ssref{assembly}).
\\
The vector of nodal coordinates of a \emph{support region} is approximated as a linear combination of the set of component modes,
\ea{
\mm q = \vector{\mm q_{\mrm b} \\ \mm q_{\Gamma} \\ \mm q_{\mrm i} } \cong \matrix{ccc}{
\mm I_{B\times B} & \mm 0 & \mm 0 \\
\mm 0 & \mm\Gamma & \mm 0 \\
\mm\Psi_{\mrm b} & \mm\Psi_{\Gamma}\mm\Gamma & \mm\Theta
}
\vector{\mm q_{\mrm b} \\ \mm \eta_\Gamma \\ \mm \eta } = \mm T \tilde{\mm q}\fp\label{eq:supportHCB}
}
Herein, the vectors $\mm q_{\mrm b}\in\mathbb R^{ B\times 1}$, $\mm q_{\Gamma}\in\mathbb R^{\Gamma\times 1}$, and $\mm q_{\mrm i}\in\mathbb R^{ I\times 1}$ contain the contact boundary, the substructure interface, and the remaining coordinates, respectively.
An appropriate sorting of $\mm q$ is assumed here for convenience.
Further, $\mm I_{ B\times  B}$ is the $ B$-dimensional identity matrix, and the vectors $\mm \eta_\Gamma\in\mathbb R^{ M_\Gamma\times 1}$, and $\mm \eta\in\mathbb R^{ M\times 1}$ contain the coordinates of the interface modes, and the fixed-interface normal modes, respectively.
Thus, the dimension of the vector of generalized coordinates of the reduced component model, $\tilde{\mm q}$, is $ B+ M_\Gamma + M$.
For a \emph{thin-walled region}, \eref{supportHCB} is analogous, only the set of contact coordinates is empty, so that the dimension of $\tilde{\mm q}$ is just $ M_{\Gamma} +  M$.
\\
The matrix $\mm\Theta$ contains as column vectors a subset of the normal modes for fixed (contact and substructure) interfaces ($\mm q_{\Gamma}=\mm 0$, $\mm q_{\mrm b}=\mm 0$), as defined in \erefs{FINMone}-\erefo{FINMtwo}.
The static constraint modes with respect to the contact interface, $\mm\Psi_{\mrm b}$, and the static constraint modes with respect to the reduced substructure interface, $\mm\Psi_{\Gamma}\mm\Gamma$, are obtained from the linear algebraic equation systems \erefo{constraintModesB}-\erefo{constraintModesG},
\ea{
\mm\Theta &=& \matrix{ccc}{\mm\theta_1 & \ldots & \mm\theta_M}\fk \label{eq:FINMone}\\
\left(\mm K_{\mrm{ii}} - \omega_j^2\mm M_{\mrm{ii}}\right)\mm\theta_j &=& \mm 0 \quad j=1,\ldots,M\fk \label{eq:FINMtwo}\\
\mm K_{\mrm{ii}}\mm\Psi_{\mrm b} &=& -\mm K_{\mrm{ib}}\fk \label{eq:constraintModesB}\\
\mm K_{\mrm{ii}}\mm\Psi_{\Gamma}\mm\Gamma &=& -\mm K_{\mrm{i}\Gamma}\mm\Gamma\fp \label{eq:constraintModesG}
}
Herein, $\mm K_{\mrm{ii}}$, $\mm K_{\mrm{ib}}$ and so on are the respective partitions of $\mm K$, and analogously for $\mm M$.
It is important to note that $\mm\Psi_{\Gamma}$ does not have to be explicitly computed, but only the product $\mm\Psi_{\Gamma}\mm\Gamma$ with the interface modes $\mm\Gamma$.
This reduces the effort for computing the corresponding static constraint modes \cite{Carassale.2017}.
\begin{figure}[ht!]
\centering
\begin{subfigure}{0.32\textwidth}
    \centering
    \includegraphics[]{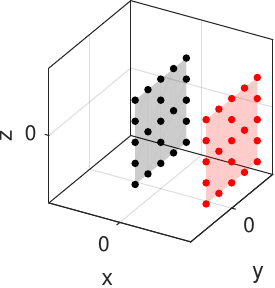}
\end{subfigure}
\hfill
\begin{subfigure}{0.32\textwidth}
    \centering
    \includegraphics[]{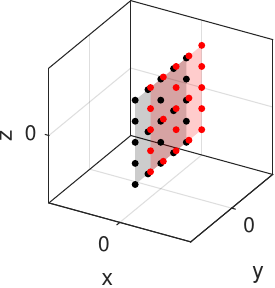}
\end{subfigure}
\hfill
\begin{subfigure}{0.32\textwidth}
    \centering
    \includegraphics[]{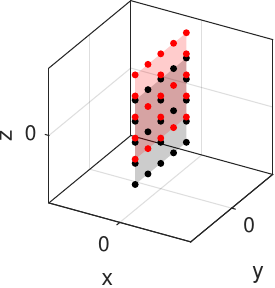}
\end{subfigure}
\begin{subfigure}{0.32\textwidth}
    \centering
    \includegraphics[]{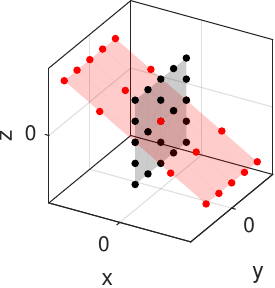}
\end{subfigure}
\hfill
\begin{subfigure}{0.32\textwidth}
    \centering
    \includegraphics[]{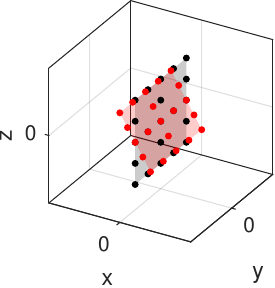}
\end{subfigure}
\hfill
\begin{subfigure}{0.32\textwidth}
    \centering
    \includegraphics[]{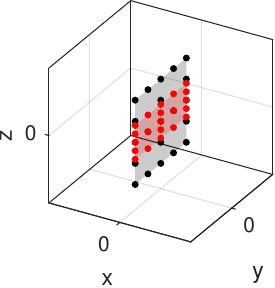}
\end{subfigure}
\begin{subfigure}{0.32\textwidth}
    \centering
    \includegraphics[]{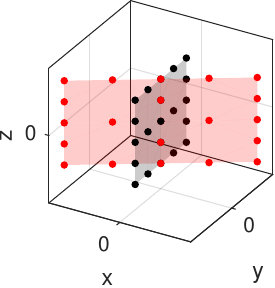}
\end{subfigure}
\hfill
\begin{subfigure}{0.32\textwidth}
    \centering
    \includegraphics[]{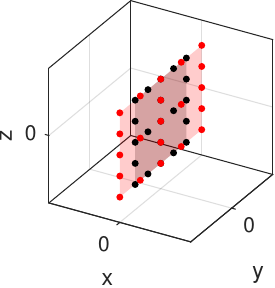}
\end{subfigure}
\hfill
\begin{subfigure}{0.32\textwidth}
    \centering
    \includegraphics[]{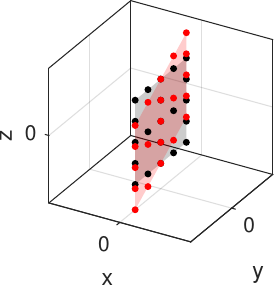}
\end{subfigure}
\caption{Illustration of an orthogonal polynomial series on a rectangular interface. First row corresponds to degree-zero, second and third row to degree-one polynomial terms.}
\label{fig:final_IMs}
\end{figure}
\\
In accordance with the interface reduction approach proposed in \cite{Carassale.2017}, the columns of $\mm\Gamma$ represent a set of orthogonal polynomial terms defined on the interface, and evaluated at the node locations, as detailed in \aref{OPS}.
For the example of a rectangular interface, the lowest-degree polynomial terms are illustrated in \fref{final_IMs}.
It is important to understand that some of these correspond to rigid-body (translation and rotation), and others to elastic displacements.
The first row shows the degree-zero polynomial terms, which correspond to translation in $x$-, $y$- and $z$-direction, respectively.
The second and third row show all degree-one polynomial terms.
The mode in the second row, first column corresponds to a rotation around the $x$-axis.
The mode in the third row, first column corresponds to a rotation around the $z$-axis.
The modes in the second row, second column, and that in the third row, third column, correspond to an in-plane shear deformation, and, properly combined, to a rotation around the $y$-axis.
The modes in the second row, third column, and that in the third row, second column, correspond to in-plane stretching in the $z$- and the $y$-direction, respectively.

\subsubsection{Support region\label{sec:supportROM}}
Substituting \eref{supportHCB} into \eref{supportFOM}, and requiring orthogonality with respect to the set of component modes, yields the reduced-order model of the support region,
\ea{
\tilde{\mm M}\ddot{\tilde{\mm q}}+ \tilde{\mm K}\tilde{\mm q} + \tilde{\mm f}_{\mrm{con}} + \tilde{\mm c} &=& \tilde{\mm f}_{\mrm{ext}}\fk\label{eq:sROM}
}
where
\ea{
\tilde{\mm M} &=& \mm T^{\mrm T}\mm M\mm T\fk \label{eq:mtil}\\
\tilde{\mm K} &=& \mm T^{\mrm T}\mm K\mm T\fk \label{eq:ktil}\\
\tilde{\mm f}_{\mrm{con}} &=& \mm T^{\mrm T}{\mm f}_{\mrm{con}}\fk\label{eq:fcontil}\\
\tilde{\mm c} &=& \mm T^{\mrm T}{\mm c}\fk\label{eq:ctil}\\
\tilde{\mm f}_{\mrm{ext}} &=& \mm T^{\mrm T}{\mm f}_{\mrm{ext}}\fp\label{eq:fexttil}
}
Herein, $\mm T$ is that defined in \eref{supportHCB}.
\\
We propose to use the relative displacements at the contact interface (\emph{contact gap}), $\mm g$, as contact boundary coordinates, $\mm q_{\mrm b}=\mm g$.
Compared to using the absolute displacements of the nodes on both sides of the contact interface, this reduces the number of static constraint modes that have to be computed, and is known to improve the convergence with respect to the number of retained normal modes.
The transform to relative displacements is valid under the assumption of small sliding, which is reasonable for jointed structures.
A contact law is needed, which locally relates the contact stress with the contact gap and its time derivatives, along with a quadrature scheme to obtain the consistent nodal forces.
The proposed method is not limited to a particular contact law, nor a particular quadrature scheme.
An example implementation is specified in \sref{numerical}.
The contact forces are collected in a vector ${\mm f}_{\mathrm{con,b}}$.
In general, the contact stress, and thus $\tilde{\mm f}_{\mathrm{con}}$, is independent of the modal coordinates $\mm\eta_\Gamma$ or $\mm \eta$, and thanks to the structure of $\mm T$ in \eref{supportHCB}, we have $\tilde{\mm f}_{\mathrm{con}} = [{\mm f}_{\mathrm{con,b}}; \mm 0; \mm 0]$.
In this sense, sparsity of the terms associated with nonlinear contact behavior is retained in the reduced component model.
This permits the block elimination of the linear part in algebraic equation systems arising \eg for implicit time stepping schemes.

\subsubsection{Thin-walled region}
The reduced component model of a thin-walled region is obtained analogous to \eref{sROM} and reads:
\ea{
\tilde{\mm M}\ddot{\tilde{\mm q}} + \tilde{\mm K}\tilde{\mm q} + \tilde{\mm f}_{\mrm{geom}} + \tilde{\mm c} &=& \tilde{\mm f}_{\mrm{ext}}\fp\label{eq:tROM}
}
While \erefs{mtil}-\erefo{ktil}, \erefo{ctil}-\erefo{fexttil} apply also to the thin-walled regions, $\tilde{\mm f}_{\mrm{geom}}$ will not be computed as a projection, $\tilde{\mm f}_{\mrm{geom}}=\mm T^{\mrm T}\mm f_{\mrm{geom}}$, as opposed to \eref{fcontil}, but using implicit condensation, as described in \ssref{IC}.

\subsection{Implicit condensation with an original engineering-oriented load scaling\label{sec:IC}}
Within the proposed sub-structuring approach, in principle, the available methods presented in \ssref{NLROM} can be used to model geometrically nonlinear components within the thin-walled regions.
More specifically, those belonging to the categories (b) and (c) are applicable, since those of category (a) are not compatible with sub-structuring.
For the present work, implicit condensation was selected.
More specifically, each element $\tilde{f}_{\mrm{geom},i}$ of the vector $\tilde{\mm f}_{\mrm{geom}}$ is approximated as
\ea{
\tilde{f}_{\mrm{geom},i} = \sum\limits_{j=1}^{R}\sum\limits_{k=j}^R \beta_{2,i}^{jk}\tilde{q}_j \tilde{q}_k + \sum\limits_{j=1}^{R}\sum\limits_{k=j}^R\sum\limits_{l=k}^{R}\beta_{3,i}^{jkl}\tilde{q}_j \tilde{q}_k \tilde{q}_l\fk \label{eq:NLgeomTAYLOR}
}
where $R$ is the number of component modes ($R=M+M_\Gamma$ for the proposed reduced model of thin-walled components).
In \eref{NLgeomTAYLOR}, the sums start in such a way that redundant polynomial terms are avoided.
As explained in the introduction, \eref{NLgeomTAYLOR} is to be interpreted as a cubic-order Taylor series expansion around $\tilde{\mm q}=\mm 0$.
Within implicit condensation, the coefficients $\lbrace\beta_{2,i}^{jk}\rbrace$, $\lbrace\beta_{3,i}^{jkl}\rbrace$ are determined by regression to a set of static load cases.
Within this subsection, the main original contribution is the proposed load scaling.
\\
The thin-walled regions are considered individually, in the absence of dynamic coupling forces and inertia forces; \ie, $\mm c^{(w)}=\mm 0=\mm M^{(w)}\ddot{\mm q}^{(w)}$ in \eref{thinwalledFOM}.
Omitting the specifier $\square^{(s)}$ this leads to
\ea{
\mm K \mm q + \mm f_{\mrm{geom}}\left(\mm q\right) = \mm K\mm T\mm w = \begin{cases}
\mm T_j w_j & \text{single-mode load case} \\
\mm T_j w_j + \mm T_k w_k & \text{two-mode load case} \\
\mm T_j w_j + \mm T_k w_k + \mm T_l w_l & \text{three-mode load case}
\end{cases}\fp\label{eq:staticProblemIC}
}
As external forces $\mm f_{\mrm{ext}}$, the linear-elastic forces $\mm K\mm T\mm w$, induced by a linear combination of the component modes, are considered.
In \eref{staticProblemIC}, $\mm T_j$ is the $j$-th column of $\mm T$, and $w_j$ denotes the load scale, which is set as explained in the following paragraphs.
Once the displacement response, $\mm q$, has been obtained for all static load cases, the polynomial coefficients in \eref{NLgeomTAYLOR} are obtained by regression.
To make the present article self-contained, this step is described in \aref{regression}.
\\
As mentioned before, the estimated polynomial coefficients are highly sensitive to the scaling of the load cases.
If the loads are too small, the estimated coefficients are distorted by numerical noise, and if they are too large, higher-order nonlinear terms (higher than the cubic order presumed in \eref{NLgeomTAYLOR}) are activated, which distorts the estimated lower-order coefficients.
In particular, high-order nonlinear terms are important if \emph{buckling} occurs.
In that case, the structure transitions from one static equilibrium to another, which can not be approximated well with a low-order Taylor series expansion around one of the equilibrium configurations.
It is important to emphasize that this is not a hypothetical issue, but within the sub-structuring framework, buckling of thin-walled regions is almost inevitable.
For instance, the interface mode depicted in the first row, first column in \fref{final_IMs}, is directly associated with membrane stretching/compression.
For this reason, it was proposed in \cite{Kuether.2017} to apply a by factor 1000 smaller load scale to such interface modes than to the remaining modes.
While this led to reasonable results, it cannot be transferred to more complicated geometries, where the modes cannot simply be categorized into bending and membrane ones.
Also, the load scale for which buckling occurs, will depend sensitively on whether a pure membrane load case is considered, or it is combined with a bending mode.
\\
The proposed load scaling scheme relies on three individual criteria:
Besides buckling avoidance, a certain maximum target displacement $q_{\mrm{ref}}$, and a stress limit $\sigma_{\mrm{lim}}$ are specified.
The algorithm for the \emph{single-mode load cases} is as follows:
\begin{enumerate}
    \item Determine the linear estimate of the scale needed to reach the target displacement:
\ea{
\hat w_j = \frac{q_{\mrm{ref}}}{\left|\left|\mm T_j\right|\right|_\infty}\fp\label{eq:scalingDisplacement}
}
    \item Determine the buckling load factor $\gamma_{\mrm{crit},j}$, with respect to the load $\mm f_{\mrm{ext}} = \mm K\mm T_j\hat w_j\gamma_{\mrm{crit}, j}$ (from a linear analysis). Then set
\ea{
\hat\gamma_j = \begin{cases}
0.5\gamma_{\mrm{crit},j} & \text{buckling occurs} \\
1 & \text{no buckling occurs}
\end{cases}\fp\label{eq:scalingBuckling}
}
    \item Successively increase the load $w_j$ in the nonlinear static analysis until $\hat w_j \hat \gamma_j$. Evaluate the maximum stress $\sigma(w_j)$ within the component and check if the stress limit is reached. Then set
\ea{
\hat\sigma_i = \begin{cases}
\frac{w_j^*}{\hat w_j\hat\gamma_i} & \text{with}\,\,  \sigma(w_j^*)=\sigma_{\mrm{lim}}\\
1 & \text{if}\,\, \sigma(\hat w_j\hat\gamma_j)\leq\sigma_{\mrm{lim}}
\end{cases}\fp\label{eq:scalingStress}
}
    \item Finally, set $w_j = \hat w_j \hat\gamma_j \hat\sigma_j$.
\end{enumerate}
A scalar stress measure is needed; the von Mises stress is used in this work.
The single-mode load cases are treated first.
The determined scales are adopted for the \emph{multi-mode load cases}.
Steps 2.-4. are then applied analogously.
\\
It is useful to note that the only user-defined input for the proposed load scaling scheme are the parameters $q_{\mrm{ref}}$ and $\sigma_{\mrm{lim}}$.
Reasonable limits for those parameters are usually available.
For instance, the yield strength can be used for $\sigma_{\mrm{lim}}$.
After the system model has been simulated, one can easily check whether the target displacement $q_{\mrm{ref}}$ was exceeded, and increase it if necessary.
Overall, this leads to a systematic, engineering-oriented load scaling scheme.

\subsection{Assembly of the reduced system model\label{sec:assembly}}
Having reduced the component models, we return to the system level.
Thus, the specifiers $\square^{(s)}$ and $\square^{(w)}$ for support and thin-walled region, omitted in \ssrefs{CMS}-\ssrefo{IC}, are used again.
With \erefs{sROM} and \erefo{tROM}, one obtains an approximation of \erefs{supportFOM}-\erefo{thinwalledFOM}:
\ea{
\tilde{\mm M}^{(s)}\ddot{\tilde{\mm q}}^{(s)}+ \tilde{\mm K}^{(s)}\tilde{\mm q}^{(s)} + \tilde{\mm f}_{\mrm{con}}^{(s)} + \tilde{\mm c}^{(s)} &=& \tilde{\mm f}_{\mrm{ext}}^{(s)}\fk\label{eq:supportROM}
\\
\tilde{\mm M}^{(w)}\ddot{\tilde{\mm q}}^{(w)} + \tilde{\mm K}^{(w)}\tilde{\mm q}^{(w)} + \tilde{\mm f}_{\mrm{geom}}^{(w)} + \tilde{\mm c}^{(w)} &=& \tilde{\mm f}_{\mrm{ext}}^{(w)}\fk\label{eq:thinwalledROM}
}
where the reduced mass and stiffness matrices, and the external force vector are given in \erefs{mtil}-\erefo{ktil}, \erefo{fexttil}.
The contact forces are described in \ssref{supportROM}, and the geometrically nonlinear terms, $\tilde{\mm f}_{\mrm{geom}}^{(w)}$, are obtained from implicit condensation (\ssref{IC}).
With this, it remains to define the coupling forces.
\\
Recall that the substructure interface is an artificial one; \ie, it goes through a part.
Thus, compatibility of displacements on both sides of the interface must hold.
Thanks to the selected method of interface reduction, this is simply ensured by requiring that the generalized coordinates associated with corresponding interface modes are the same.
To illustrate this, consider the example of one support region ($\square^{(1)}$), and one thin-walled region ($\square^{(2)}$).
The compatibility condition then reads $\mm\eta^{(1)}_\Gamma = \mm\eta^{(2)}_{\Gamma}$.
It is proposed to ensure this condition via \emph{primal assembly}.
Therefore, a unique set of coordinates, $\mm\eta_\Gamma^{(1,2)}$, is introduced on system level, which is assigned to the respective coordinates on component level, $\mm\eta^{(1)}_\Gamma=\mm\eta_\Gamma^{(1,2)}$, and $\mm\eta^{(2)}_\Gamma=\mm\eta_\Gamma^{(1,2)}$.
This can be expressed using the matrix $\mm L$:
\ea{
\vector{
\mm q_{\mrm b}^{(1)} \\
\mm\eta_\Gamma^{(1)} \\
\mm\eta^{(1)} \\
\mm\eta_\Gamma^{(2)} \\
\mm\eta^{(2)}
}
=
\underbrace{\matrix{cccc}{
\mm I & \mm 0 & \mm 0 & \mm 0\\
\mm 0 & \mm I & \mm 0 & \mm 0\\
\mm 0 & \mm 0 & \mm I & \mm 0\\
\mm 0 & \mm I & \mm 0 & \mm 0\\
\mm 0 & \mm 0 & \mm 0 & \mm I
}}_{\mm L}
\underbrace{\vector{
\mm q_{\mrm b}^{(1)} \\
\mm\eta_\Gamma^{(1,2)} \\
\mm\eta^{(1)} \\
\mm\eta^{(2)}
}}_{\tilde{\mm q}}\fp \label{eq:compatibility}
}
Besides compatible displacement, force equilibrium must hold across the interface.
This can be shown to lead to $\mm L\tra \vector{\tilde{\mm c}^{(1)}\\ \tilde{\mm c}^{(2)} } = \mm 0$ \cite{Allen.2020}.
More specifically, this condition follows from the fact that the reaction forces, $\tilde{\mm c}^{(1)}$ and $\tilde{\mm c}^{(2)}$, needed to ensure the constraint defined in \eref{compatibility}, must not produce any virtual work on system level.
\\
By projecting the decoupled equations \erefo{supportROM}-\erefo{thinwalledROM} onto the matrix $\mm L$, one eliminates the reaction forces, and obtains the reduced system model:
\ea{
\tilde{\mm M}\ddot{\tilde{\mm q}} + \tilde{\mm K}\tilde{\mm q} + \tilde{\mm h} = \tilde{\mm f}_{\mrm{ext}}\fk\label{eq:systemROM}
}
where
\ea{
\tilde{\mm M} &=& \mm L\tra \matrix{cc}{\tilde{\mm M}^{(1)} & \mm 0\\ \mm 0 & \tilde{\mm M}^{(2)}} \mm L\fk \quad 
\tilde{\mm K} = \mm L\tra \matrix{cc}{\tilde{\mm K}^{(1)} & \mm 0\\ \mm 0 & \tilde{\mm K}^{(2)}} \mm L\fk \\ 
\tilde{\mm h} &=& \vector{{\mm f}_{\mrm{con,b}}^{(1)} \\ \tilde{\mm f}_{\mrm{geom},\Gamma}^{(2)} \\ \mm 0 \\ \tilde{\mm f}_{\mrm{geom,i}}^{(2)} }\fk\quad 
\tilde{\mm f}_{\mrm{ext}} = \mm L\tra \vector{\tilde{\mm f}_{\mrm{ext}}^{(1)} \\ \tilde{\mm f}_{\mrm{ext}}^{(2)}}\fp 
}
The final dimension of the reduced system model is $ B +  M_\Gamma +  M^{(1)} +  M^{(2)}$.

\subsection{Overview of proposed method; practical implementation aspects\label{sec:overview}}
An overview of the proposed method for model reduction of thin-walled jointed structures is given in \fref{alg_substructuring}.
\begin{figure}[ht!]
    \centering
    \includegraphics{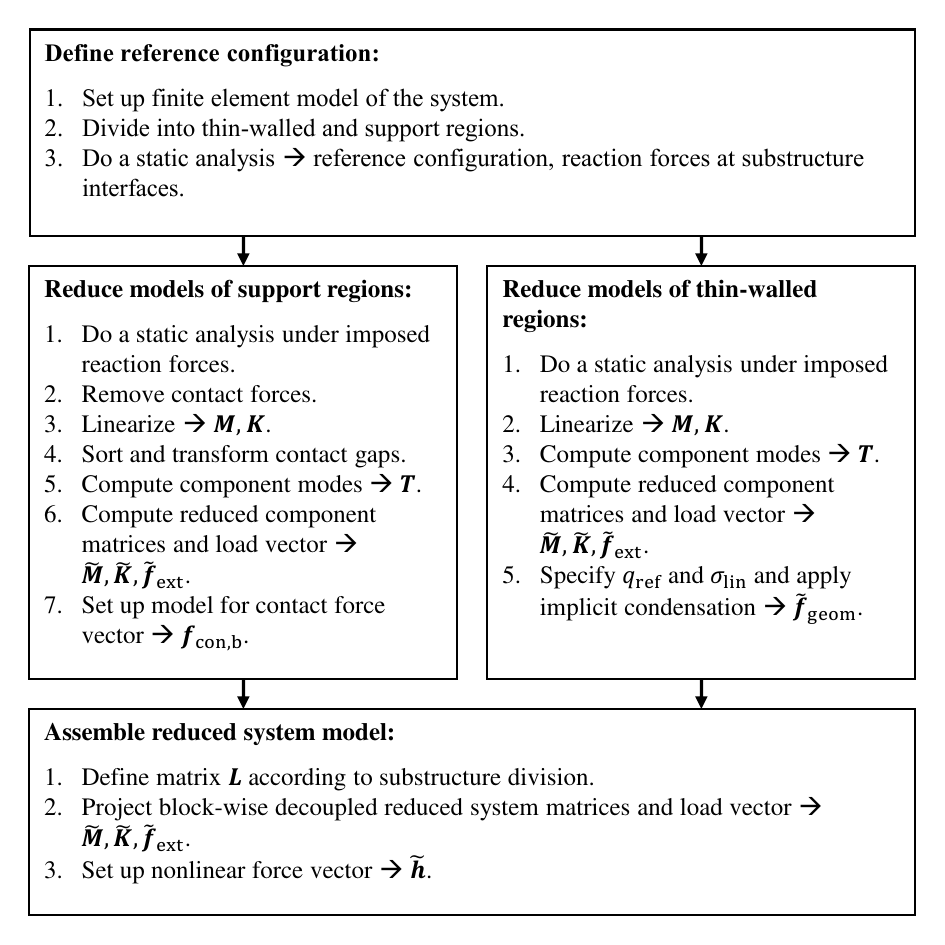}
    \caption{Overview of the proposed method.}
    \label{fig:alg_substructuring}
\end{figure}
\\
Most finite element tools can be used to carry out the tasks required for the proposed method (meshing of 3D geometries; static analysis under consideration of geometrically nonlinear and nonlinear contact behavior; output of reaction forces, nodal displacements and stresses as result of imposed forces and displacement constraints specified in a file; linear perturbation and export of tangent stiffness and mass matrices with associated degree-of-freedom maps).
In the present work, we used \ABAQUS for those tasks.
The remaining tasks of the proposed method (coordinate transforms; linear static and eigenvalue analyses; inner products of vectors and matrices; exporting/importing text files for load definition and response acquisition; regression using pseudo-inverse; solution of nonlinear algebraic governing equations; time step integration) can be coded in many programming languages with state-of-the-art linear algebra capabilities.
In the present work, we used \MATLAB for those tasks.
In the following, a few practical aspects of the implementation are mentioned for completeness.
\\
The implementation of \eref{staticProblemIC} is slightly more involved than it seems on first sight:
Recall that $\mm q$ counts from the reference configuration, \ie, the static equilibrium of the \emph{assembly}.
However, the thin-walled component must be considered in its uncoupled configuration in \eref{staticProblemIC}, because the substructure interface degrees of freedom must be available.
To resolve this, the reaction forces at the substructure interface are determined from the static analysis of the assembly, and they are imposed on the decoupled thin-walled region in a second step.
This ensures that the thin-walled region is in its reference configuration, and the nodes at the substructure interface are not constrained.
Subsequently, the load cases according to \eref{staticProblemIC} are applied.
\\
As the thin-walled region is considered in its uncoupled configuration, it typically has rigid-body degrees of freedom.
All possible rigid-body degrees of freedom (three translations, three rotations), are described by the degree-zero and degree-one polynomial terms illustrated in \fref{final_IMs}.
Of course, rigid body motion does not produce any strains; the associated interface modes are to be excluded from $\tilde{\mm f}_{\mrm{geom}}(\tilde{\mm q})$, and the corresponding polynomial coefficients are to be set to zero.
In the presence of rigid-body degrees of freedom, special attention is required to ensure that the static problem is well-posed.
\\
It was found useful to make the component modes (columns of matrix $\mm T$) dimensionless.
The static constraint modes in the first hyper-column of \eref{supportHCB} are dimensionless thanks to the identity matrix.
Further, the individual polynomial terms were normalized so that each column of $\mm\Gamma$ has maximum absolute entry of unity.
Finally, the fixed-interface normal modes were normalized by the condition $\mm\theta_j^{\mrm T}\mm M_{\mrm{ii}}/\mrm{kg}\mm\theta_j=1$; \ie, they are mass-normalized without inheriting the unit one over square-root of mass.
\\
It is useful to recall at this point the main parameters of the proposed method that have to be selected by the user.
In each support region, the number of normal modes, $M^{(j)}$, of each component, and the number of interface modes, $M_{\Gamma}$, of each interface are to be set.
It is common practice to specify a \emph{target frequency band}, which spans the relevant spectrum of external and internal forces, and retain those normal modes that have their natural frequency in that band.
In the presence of symmetries, one may be able to filter out some normal and interface modes that are not expected to contribute to the response.
However, in general, the numbers of normal and interface modes will have to be selected based on experience.
To gain confidence, one may increase those numbers in an iterative manner until the results stabilize.
\\
The implicit condensation is the bottleneck of the proposed method.
The required computation effort increases rapidly with the number of interface modes.
Therefore, any means to reduce the required number of interface modes should be considered.
The smaller the interface, and the simpler its kinematics, the fewer interface modes are expected to be sufficient.
To some extent, this can be influenced by the choice of the precise location of the interface between thin-walled and support region.
Ideally, the interface is far from any source non-uniformity, so that rigid-body-type interface modes are sufficient.
However, the interface should not be too far into the thin-walled region because otherwise some parts of the geometrically nonlinear behavior could be lost.

\section{Numerical Validation and performance assessment\label{sec:numerical}}
%
\begin{figure}[ht!]
\centering
    \begin{subfigure}[b]{.45\textwidth}
    \centering
    \includegraphics[width=\textwidth]{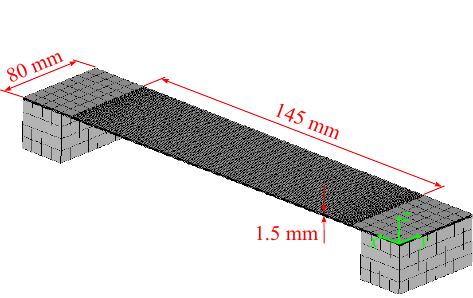}
    \caption{}
    \label{fig:full_model}
    \end{subfigure}
    \hfill
    \begin{subfigure}[b]{.45\textwidth}
    \centering
    \includegraphics[width=\textwidth]{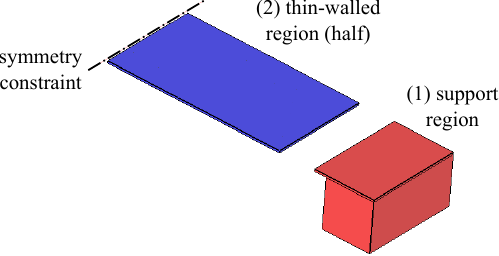}
    \caption{}
    \label{fig:substructures}
    \end{subfigure}
    \begin{subfigure}[b]{.45\textwidth}
    \centering
    \includegraphics[width=\textwidth]{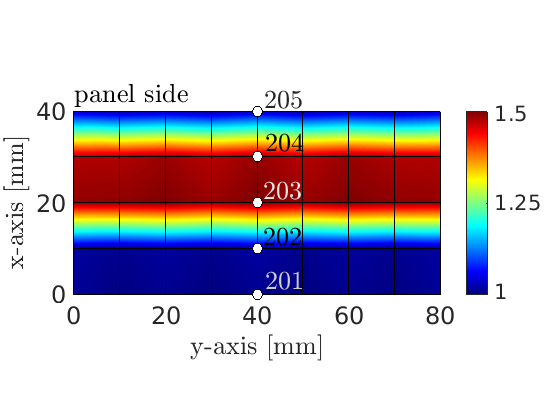}
    \caption{}
    \label{fig:CPRESS}
    \end{subfigure}
    \hfill
    \begin{subfigure}[b]{.45\textwidth}
    \centering
    \includegraphics[width=\textwidth]{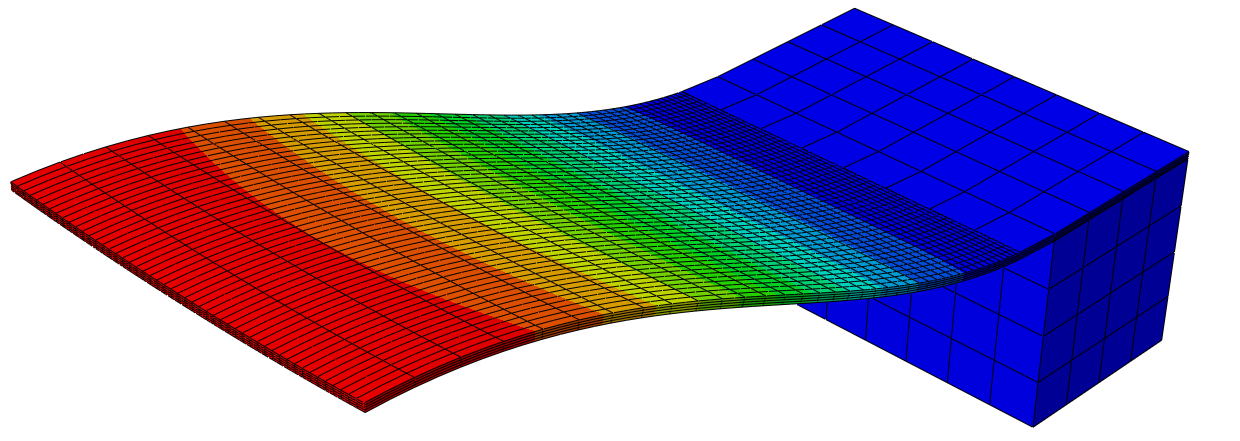}
    \caption{}
    \label{fig:lin_modeshape}
    \end{subfigure}
    \caption{Benchmark problem consisting of a panel with frictional clamping: (a) finite element model; (b) division of symmetric half of system into thin-walled (blue) and support (red) region; (c) normalized initial contact pressure distribution $\chi(x,y)$; (d) deflection shape of lowest-frequency bending mode of underlying linear system
    }
    \label{fig:FEmodel}
\end{figure}
%
In this work, a plate (or panel) with frictional clamping was considered, as illustrated in \fref{FEmodel}.
The dimensions and material parameters are similar to the benchmark system of the Tribomechadynamics Research Challenge \cite{TRCprediction,Muller.2022c}.
Both the structure and the considered load scenarios are symmetric.
This was exploited by modeling just the symmetric half and imposing appropriate constraints.
More specifically, all nodal displacements in $x$-direction are zero at the symmetry interface.
The dimensions of the problem are given in \fref{FEmodel}a.
A homogeneous, linear-elastic, isotropic material was considered (Young's modulus $E=207~\mrm{GPa}$, Poisson ratio $0.288$, mass density $7,829~\mrm{kg}/\mrm{m}^3$), both for the panel and for the support block.
The system was divided into support region ($\square^{(1)}$), and thin-walled (half) region ($\square^{(2)}$), as indicated in \fref{FEmodel}b.
The finite element model of the thin-walled region contains $8,000$ hexahedral elements with
quadratic shape functions. 
To avoid hourglassing, elements with full integration have been used.
Further, to avoid artificial stress localizations, the mesh of the panel was progressively refined towards the substructure interface.
The substructure interface has $4$ quadrilateral elements in the thickness ($z$), and $50$ in the width ($y$) direction.
The finite element model of the support region contains $288$ hexahedral elements with linear shape functions. 
The interface between support and thin-walled region was placed one element row behind the end of the support block; \ie, it extends one element row into the panel.
One motivation for this was the strategy to prevent rigid-body motion during the static analysis needed for implicit condensation, as detailed below.
Recall also the general thoughts about the precise location of the interface from the last paragraph in \ssref{overview}.
Based on the results of the present work, we do not expect a high sensitivity of the proposed method with respect to the precise location of the interface; still a systematic investigation could be useful future work.
\\
For implicit condensation, a target displacement of $q_{\mrm{ref}}=3~\mrm{mm}$, corresponding to twice the panel thickness, and a stress limit of $\sigma_{\mrm{lim}}=500~\mrm{MPa}$ were specified.
The symmetry constraint excludes rigid-body translation in $x$-direction, and rotation around the $z$- and the $y$-direction.
Rigid-body translation in $y$-direction was treated using inertia relief.
Rigid-body translation in $z$-direction, and rotation around the $x$-direction were excluded by constraining the $z$-displacement of the nodes in the second row from the substructure interface, top surface.
Those artificial constraints were applied because using inertia relief alone led to convergence failure within the static analyses.
The accuracy of the reduced model of the thin-walled region obtained with implicit condensation was thoroughly assessed.
In particular, the usefulness and benefits of the proposed load scaling were analyzed.
Representative results are presented in \aref{ICassessment}.
\\
Contact was considered on the flat rectangular interface between panel and support block.
The preloading mechanism (\eg using bolts) was not modeled.
Instead, an initial pressure distribution was simply prescribed, $p_{\mrm n0} = \overline{p}_{\mrm{n}0}\chi(x,y)$.
Three different levels $\overline{p}_{\mrm{n}0}$ were considered, as specified later, while the same non-homogeneous distribution $\chi(x,y)$ illustrated in \fref{FEmodel}c was considered in all cases.
In the contact normal direction, a unilateral-elastic constitutive relation between pressure $p_{\mrm n}$ and normal gap $g_{\mrm{n}}$ was considered,
\ea{
p_{\mrm n} = \max\left(p_{\mrm n0} + k_{\mrm n} g_{\mrm n},0\right)\fk \label{eq:unilateralSpring}
}
where $k_{\mrm n}$ is the normal stiffness per area, set to $k_{\mrm n}=10^4~\mrm{N}/\mrm{mm}^3$, which is of a realistic order of magnitude for the given scenario \cite{lacayo2019}.
In the tangential direction, an elastic dry friction law was considered.
The relation between traction vector $\mm p_{\mrm t}$ and tangential gap $\mm g_{\mrm t}$ is governed by the incremental evolution law
\ea{
\Delta \mm p_{\mrm{t}} = \begin{cases}
k_{\mrm t}\Delta \mm g_{\mrm t} & \text{(sticking)}\,\, \|\mm p_{\mrm{t}} + k_{\mrm t}\Delta \mm g_{\mrm t}\| \leq \mu p_{\mrm n} \\
\mu p_{\mrm n} \frac{\Delta \mm g_{\mrm t}}{\|\Delta \mm g_{\mrm t}\|} - \mm p_{\mrm t} & \text{(sliding)}\,\, \|\mm p_{\mrm{t}} + k_{\mrm t}\Delta \mm g_{\mrm t}\| \geq \mu p_{\mrm n}
\end{cases}\fp
}
The increment $\Delta$ refers to the difference between two consecutive time levels.
The friction coefficient was set to $\mu =0.3$.
The tangential contact stiffness $k_{\mrm t}$ was defined in such a way that a prescribed limit stick distance $g_{\mrm{sl}}=0.1~\mum$ was obtained, where
$k_{\mrm t}=\mu p_{\mrm n}/g_{\mrm{sl}}$.
The contact meshes are conform on both sides of the interface, and have $5$ nodes in the $x$- and $9$ nodes in the $y$-direction (\fref{FEmodel}c).
A node-based integration of the contact stress was pursued, using the $ C = 4\times5$ nodes on one side of the contact interface as quadrature points, and the respective area as weight.
\\
The reduction basis is described and illustrated in \ssref{reductionBasis}, where also the convergence of the linear modal frequencies with the number of component modes is analyzed.
The amplitude-dependence of the frequency and damping ratio of the fundamental bending mode is analyzed in \ssref{NMA}.
Besides the validation of the proposed method, the relative importance of geometrically nonlinear and contact behavior, and their mutual interaction, is of particular interest here.
In \ssref{impulse}, an impulsive loading is considered, where multiple modes contribute to the nonlinear response.
An important difference of the benchmark problem in \fref{FEmodel} and the TRC benchmark system in \cite{TRCprediction} is the idealization of the support structure (no bolts nor bore holes).
This simplifies the interpretation of the response.
Also, this permits a rather coarse contact mesh, which greatly reduces the effort for the finite element reference analyses.
To demonstrate the suitability of the proposed approach for more complicated scenarios, and to illustrate its modular character, the model of the support region is replaced in \ssref{replace} by a refined model with bolts and bore holes.

\subsection{Reduction basis\label{sec:reductionBasis}}
\begin{figure}[ht!]
    \centering
    \begin{subfigure}[b]{.45\textwidth}
    \centering
    \includegraphics[width=\textwidth]{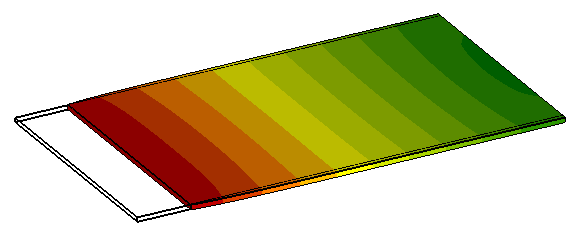}
    \caption{}
    \label{fig:infc1}
    \end{subfigure}
    \hfill
    \begin{subfigure}[b]{.45\textwidth}
    \centering
    \includegraphics[width=\textwidth]{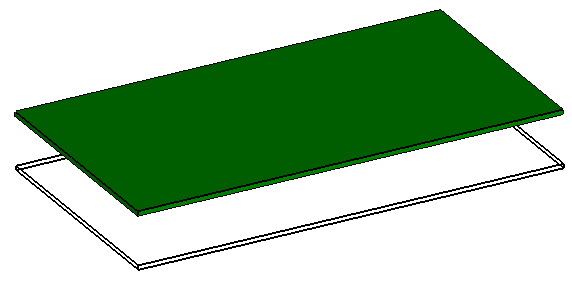}
    \caption{}
    \label{fig:infc2}
    \end{subfigure}
    \hfill
    \begin{subfigure}[b]{.45\textwidth}
    \centering
    \includegraphics[width=\textwidth]{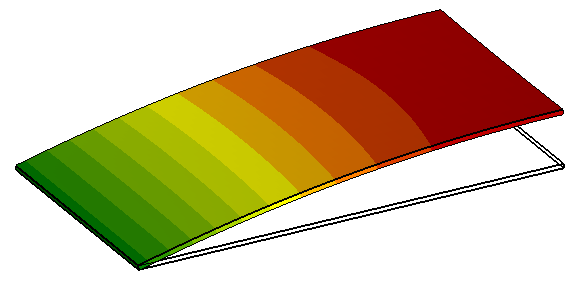}
    \caption{}
    \label{fig:infc3}
    \end{subfigure}
    \hfill
    \begin{subfigure}[b]{.45\textwidth}
    \centering
    \includegraphics[width=\textwidth]{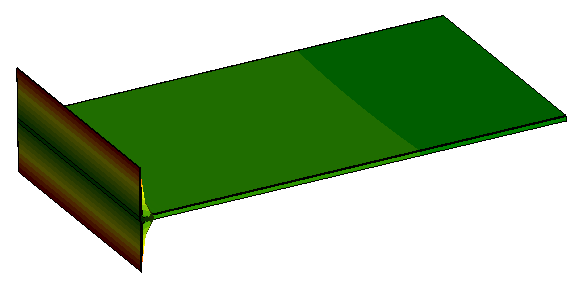}
    \caption{}
    \label{fig:infc4}
    \end{subfigure}
    \hfill
    \begin{subfigure}[b]{.45\textwidth}
    \centering
    \includegraphics[width=\textwidth]{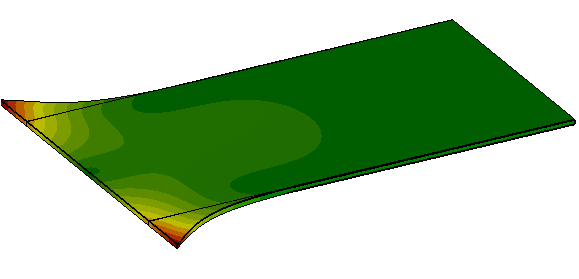}
    \caption{}
    \label{fig:infc5}
    \end{subfigure}
    \hfill
    \caption{Reduced basis: Panel interface (constraint) modes.
    The modes were normalized, respectively, so that the maximum displacement is equal among all modes, and indicated by red color, while green corresponds to zero displacement.
    }
    \label{fig:infc}
\end{figure}
%
For the interface modes, all degree-zero and degree-one polynomial terms were considered.
As the substructure interface has a rectangular cross section, the polynomial terms resemble those illustrated in \fref{final_IMs}.
Some polynomial terms were excluded due to symmetry.
This applies to the rigid-body-translation in the $y$-direction (first row, second column in \fref{final_IMs}), the
shear modes in the $y$-$z$ plane (second row, second column and third row, third column in \fref{final_IMs}), and the rigid-body-rotation around the $z$-axis (third row, first column in \fref{final_IMs}).
Excluding those four polynomial terms from the nine depicted in \fref{final_IMs}, five remain.
The corresponding interface constraint modes are those depicted in \fref{infc}.
\\
The mode in \fref{infc}(a) corresponds to an axial translation of the interface, which is important for modeling finite axial clamping stiffness, and its effect on the membrane deformation.
The mode in \fref{infc}(b) corresponds to a rigid-body translation in the lateral direction, not only of the interface, but of the whole panel.
This is important for modeling finite lateral stiffness of the support (including the small free section and the clamped section of the panel).
The mode in \fref{infc}(c) corresponds to a rotation of the interface around the bending axis, which is important for modeling finite rotational clamping stiffness.
This is expected to have a crucial effect on the bending stiffness already in the linear case.
The modes in \fref{infc}(d)-(e) correspond to elastic interface deformation, namely some tapering in the $z$- and $y$-direction, respectively.
Compared to the other modes, those constraint modes are rather localized to the domain near the interface.
As proposed in \ssref{IC}, the rigid-body mode depicted in \fref{infc}(b) was excluded from $\tilde{\mm f}_{\mrm{geom}}(\tilde{\mm q})$.
\begin{figure}[ht!]
    \centering
    \begin{subfigure}[b]{.45\textwidth}
    \centering
    \includegraphics[width=\textwidth]{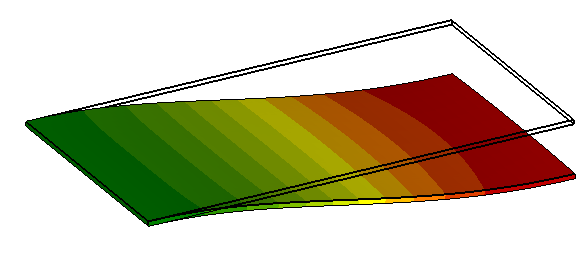}
    \caption{}
    \label{fig:fim1}
    \end{subfigure}
    \hfill
    \begin{subfigure}[b]{.45\textwidth}
    \centering
    \includegraphics[width=\textwidth]{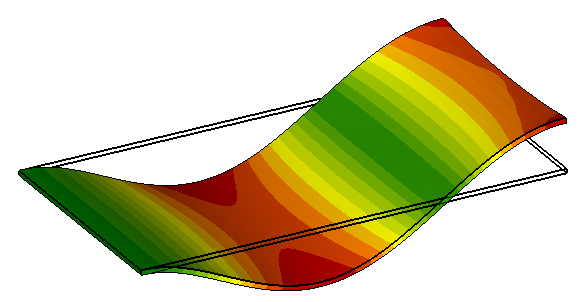}
    \caption{}
    \label{fig:fim2}
    \end{subfigure}
    \hfill
    \begin{subfigure}[b]{.45\textwidth}
    \centering
    \includegraphics[width=\textwidth]{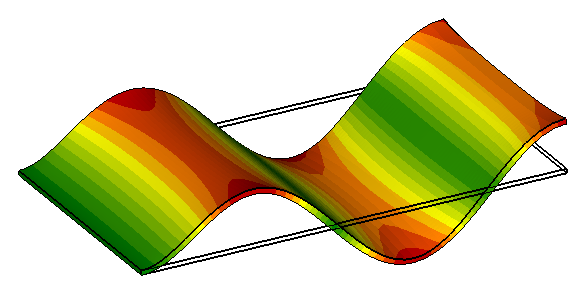}
    \caption{}
    \label{fig:fim3}
    \end{subfigure}
    \hfill
    \begin{subfigure}[b]{.45\textwidth}
    \centering
    \includegraphics[width=\textwidth]{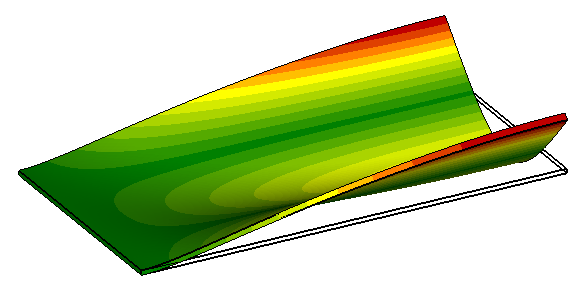}
    \caption{}
    \label{fig:fim4}
    \end{subfigure}
    \hfill
    \begin{subfigure}[b]{.45\textwidth}
    \centering
    \includegraphics[width=\textwidth]{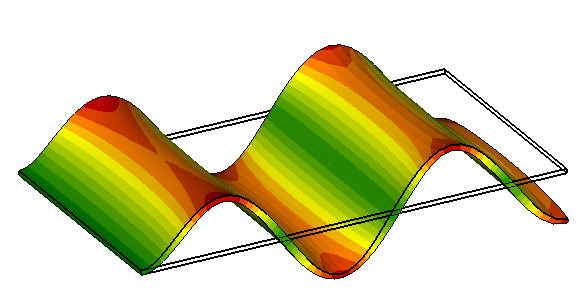}
    \caption{}
    \label{fig:fim5}
    \end{subfigure}
    \caption{Reduced basis: Panel (fixed-interface) normal modes.
    The modes were normalized, respectively, so that the maximum displacement is equal among all modes, and indicated by red color, while green corresponds to zero displacement.
    }
    \label{fig:fims}
\end{figure}
\\
Torsion modes of the panel were excluded, again, due to symmetry.
The remaining five lowest-frequency modes for fixed substructure interface are depicted in \fref{fims}.
These can be identified as the first- through forth-order bending modes in the length direction (\fref{fims}a-c,e), and the first-order bending mode in width direction (\fref{fims}d).
The modes span the frequency band from $96~\mrm{Hz}$ to $2.4~\mrm{kHz}$.
Up to $ M^{(2)}=5$ fixed-interface normal modes (depicted in \fref{fims}), and up to $ M_\Gamma=5$ interface constraint modes (depicted in \fref{infc}) were retained in the reduced model of the thin-walled region.
In the reduced model of the support region, besides the $ M_\Gamma=5$ plus $ B=3 C = 3\times4\times5$ constraint modes, $ M^{(1)}=30$ fixed-interface normal modes were retained.
The latter spanned the frequency band from $10~\mrm{kHz}$ to $17~\mrm{kHz}$.
With this, the maximum dimension of the reduced models is 170 (support region), 10 (thin-walled region), and 175 (system).
%
\begin{figure}[ht!]
    \centering
    \begin{subfigure}[b]{.45\textwidth}
    \centering
    \includegraphics[width=\textwidth]{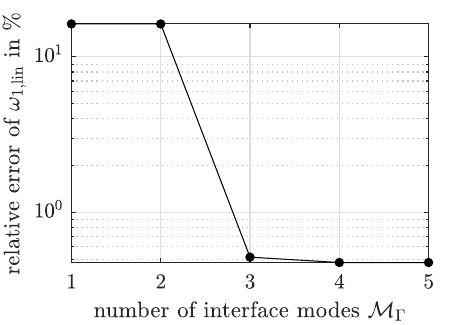}
    \caption{}
    \label{fig:lin_ims}
    \end{subfigure}
    \hfill
    \begin{subfigure}[b]{.45\textwidth}
    \centering
    \includegraphics[width=\textwidth]{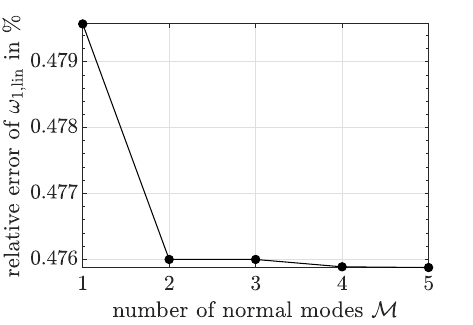}
    \caption{}
    \label{fig:rom_conv}
    \end{subfigure}
    \hfill
    \caption{
    Convergence of lowest modal frequency of the linear system:
    (a) relative error vs. $ M_\Gamma$, $ M=5$; (b) relative error vs. $ M$, $ M_\Gamma=5$.
    }
    \label{fig:linconvergence}
\end{figure}
\tab[ht]{ccccc}{
$k$ & reference & $ M_\Gamma=3, M=1$ & $ M_\Gamma=3, M=3$ & $ M_\Gamma=5, M=5$\\
1 & $82.5~\mrm{Hz}$ & $82.94~\mrm{Hz}$ ($+0.5\%$) & $82.94~\mrm{Hz}$ ($+0.5\%$)& $82.91~\mrm{Hz}$ ($+0.47\%$)\\
2 & $447.3~\mrm{Hz}$ & $-$ & $449.5~\mrm{Hz}$ ($+0.5\%$)& $448.9~\mrm{Hz}$ ($+0.36\%$)\\
3 & $1113.5~\mrm{Hz}$ & $-$ & $-$& $1117.2~\mrm{Hz}$ ($+0.33\%$) \\
4 & $1371.8~\mrm{Hz}$ & $-$& $1373.3~\mrm{Hz}$ ($+0.11\%$) & $1373.3~\mrm{Hz}$ ($+0.11\%$)\\
5 & $2084.7~\mrm{Hz}$ & $-$ & $-$ & $2101.9~\mrm{Hz}$ ($+0.82\%$)
}{Modal frequencies $\omega_{k,\mrm{lin}}/(2\pi)$ of the linear system.
}{linfreq}
\\
A linear modal analysis (tied contact) was carried out for different combinations of $ M^{(2)}$ and $ M_\Gamma$.
Results are shown in \fref{linconvergence} and listed in \tref{linfreq}.
As reference, the results obtained with the (full-order) finite element model were used.
Here and in the following, only the number of normal modes, $ M^{(2)}$, within the thin-walled region was varied, while $ M^{(1)}$ was kept constant.
For brevity, the abbreviation $ M= M^{(2)}$ is used.
For $M=1$, only the mode in \fref{fims}a is retained, for $M=2$ those in \fref{fims}a-b, for $M=3$ those in \fref{fims}a-b,d, for $M=4$ those in \fref{fims}a-d, and for $M=5$ all modes in \fref{fims}.
The error with respect to the frequency of the fundamental bending mode drops from $>16\%$ down to $<0.5\%$ when increasing $ M_\Gamma$ from $2$ to $3$.
Apparently, the constraint mode associated with interface rotation around the bending axis (\fref{infc}(c)) is more important than the constraint mode associated with lateral translation.
Further analysis showed that second-degree polynomial terms are needed to reduce the error by another order of magnitude.
Increasing the number of normal modes, $ M$, has only a relatively small influence in the linear case (\fref{linconvergence}b).
Apparently, the remaining error (for $ M_\Gamma=5= M$) is dominated by the interface modes.

\subsection{Nonlinear modal analysis; role and interaction of geometrically nonlinear and contact behavior\label{sec:NMA}}
In this subsection, the amplitude-dependence of the frequency and damping ratio of the fundamental bending mode is studied.
The corresponding modal deflection shape is illustrated in \fref{FEmodel}d.
In accordance with single-nonlinear-mode theory, those amplitude-dependent properties can be used, among others, to accurately predict the response to harmonic loading in the frequency range near a well-separated primary resonance, see \eg \cite{Krack.2015a}.
Besides assessing the accuracy of the proposed method, an important aim of the present section is to analyze the the relative importance of geometrically nonlinear and contact behavior, and their mutual interaction.
To determine the amplitude-dependent modal properties, quasi-static modal analysis was employed.
When the modal deflection shape changes significantly, \eg, when a contact interface transitions from sticking to gross slip, this method is known to yield slightly inaccurate results.
The numerical validation of the proposed model reduction method is not affected by this shortcoming, as the same modal analysis method was applied to both, the finite element and the reduced model.
The employed variant of quasi-static modal analysis is described in \aref{QSMA}.
\\
Three different initial pressure levels were considered:
$\overline{p}_{\mrm{n}0}\to\infty$ (tied contact),
$\overline{p}_{\mrm{n}0}=1.2~\mrm{MPa}$ (high initial pressure),
$\overline{p}_{\mrm{n}0}=0.8~\mrm{MPa}$ (low initial pressure).
For $\overline{p}_{\mrm{n}0}\to\infty$, the behavior of the contact is strictly linear.
This case is particularly useful to assess the accuracy of the implicit condensation.
Analogously, analyses were also carried out where only geometrically linear behavior was considered.
Following the results obtained in \ssref{reductionBasis}, four reduced models were considered, where $ M_{\Gamma}$ was set to either $3$ or $5$, and $ M$ was set to $1$, $3$ or $5$.
Recall that $ M_{\Gamma}<3$ does not lead to a reasonable approximation, already in the linear case.
As will be shown, more component modes were not necessary to achieve converged linear and nonlinear results.
It should also be mentioned that more than $10$ component modes would lead to rather high computational effort (\cf discussion at the end of \ssref{impulse}).
\begin{figure}[ht!]
    \centering
    \begin{subfigure}[b]{.45\textwidth}
    \centering
    \includegraphics[width=\textwidth]{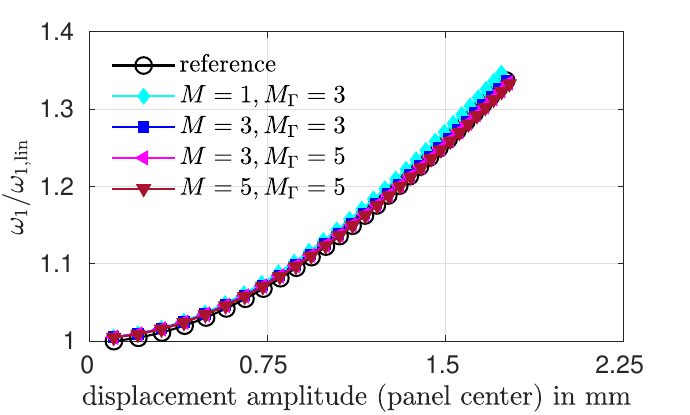}
    \caption{}
    \label{fig:omega_NLGEOM}
    \end{subfigure}
    \hfill
    \begin{subfigure}[b]{.45\textwidth}
    \centering
    \includegraphics[width=\textwidth]{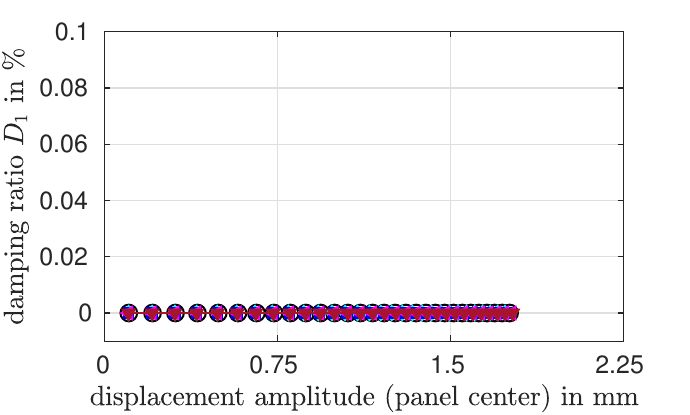}
    \caption{}
    \label{fig:D_NLGEOM}
    \end{subfigure}
    \begin{subfigure}[b]{.45\textwidth}
    \centering
    \includegraphics[width=\textwidth]{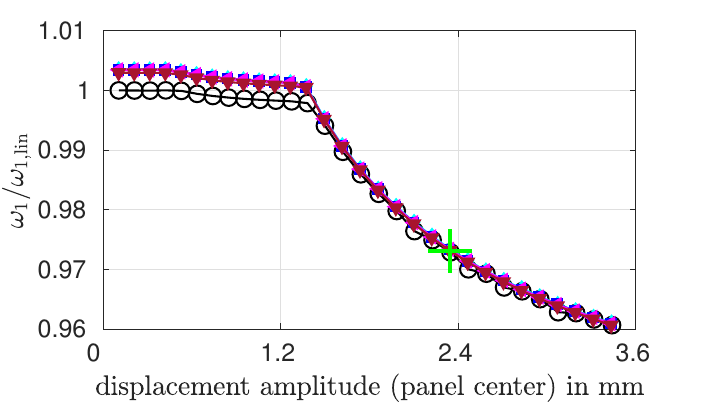}
    \caption{}
    \label{fig:omega_LINGEOM}
    \end{subfigure}
    \hfill
    \begin{subfigure}[b]{.45\textwidth}
    \centering
    \includegraphics[width=\textwidth]{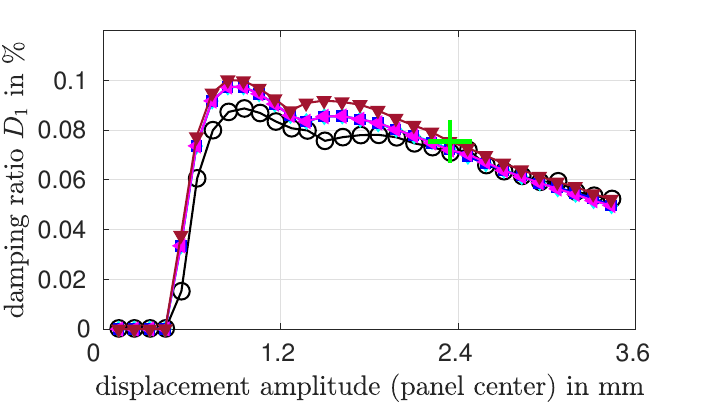}
    \caption{}
    \label{fig:D_LINGEOM}
    \end{subfigure}
    \begin{subfigure}[b]{.45\textwidth}
    \centering
    \includegraphics[width=\textwidth]{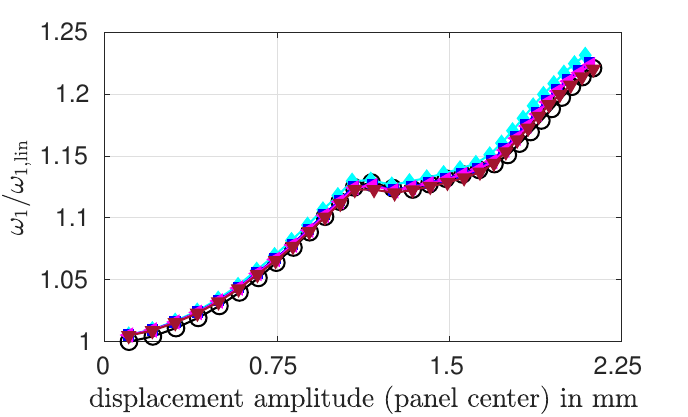}
    \caption{}
    \label{fig:omega_p1200}
    \end{subfigure}
    \hfill
    \begin{subfigure}[b]{.45\textwidth}
    \centering
    \includegraphics[width=\textwidth]{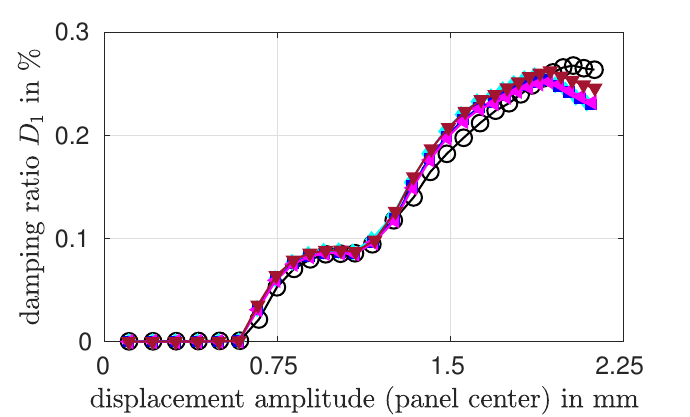}
    \caption{}
    \label{fig:D_p1200}
    \end{subfigure}
    \begin{subfigure}[b]{.45\textwidth}
    \centering
    \includegraphics[width=\textwidth]{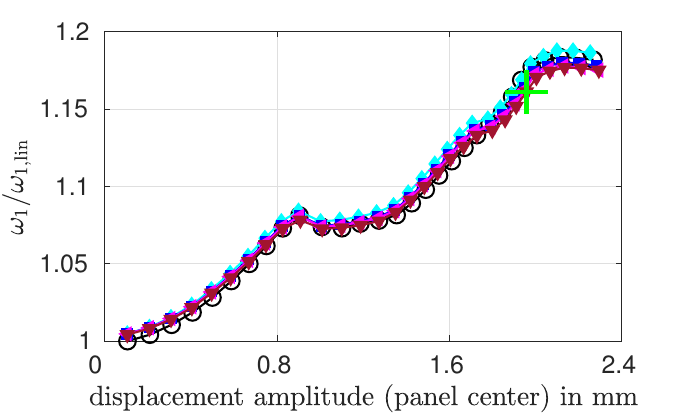}
    \caption{}
    \label{fig:omega_p800}
    \end{subfigure}
    \hfill
    \begin{subfigure}[b]{.45\textwidth}
    \centering
    \includegraphics[width=\textwidth]{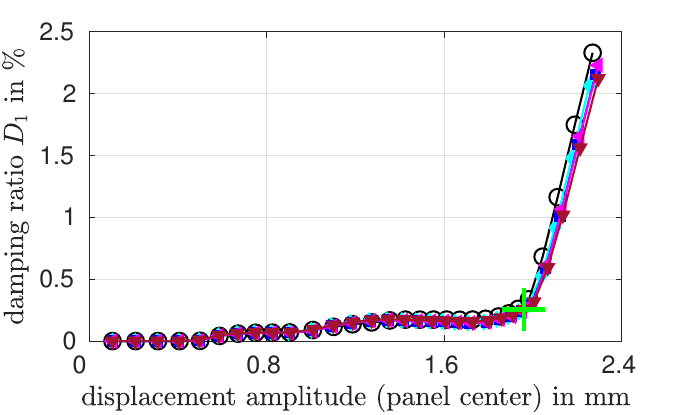}
    \caption{}
    \label{fig:D_p800}
    \end{subfigure}
    \hfill
    \caption{Amplitude-dependent modal frequency and damping ratio, respectively: (a,b) geometrically nonlinear, tied contact; (c,d) geometrically linear, low initial pressure; (e,f) geometrically nonlinear, high initial pressure; (g,h) geometrically nonlinear, low initial pressure.
    Green $+$-markers indicate points for which the contact behavior is illustrated in \fref{cstates}.
    }
    \label{fig:natural_frequencies}
\end{figure}
%
%
\begin{figure}[ht!]
    \centering
    \begin{subfigure}[b]{.45\textwidth}
    \centering
    \includegraphics[width=\textwidth]{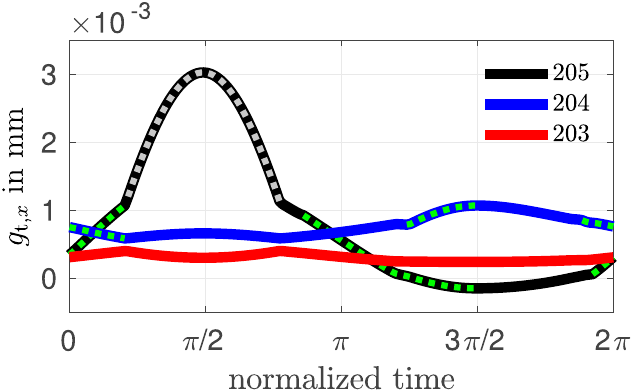}
    \caption{}
    \end{subfigure}
    \hfill
    \begin{subfigure}[b]{.45\textwidth}
    \centering
    \includegraphics[width=\textwidth]{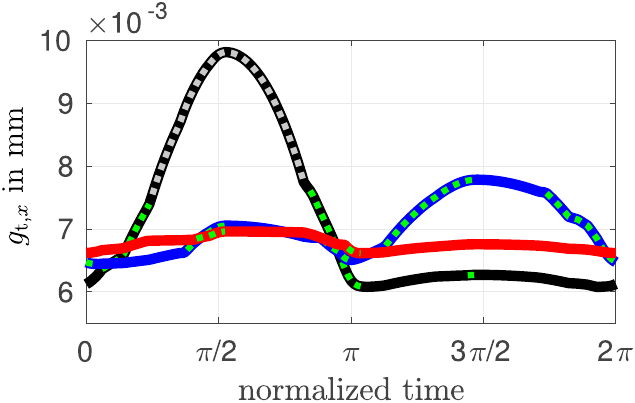}
    \caption{}
    \end{subfigure}
    \hfill
    \begin{subfigure}[b]{.45\textwidth}
    \centering
    \includegraphics[width=\textwidth]{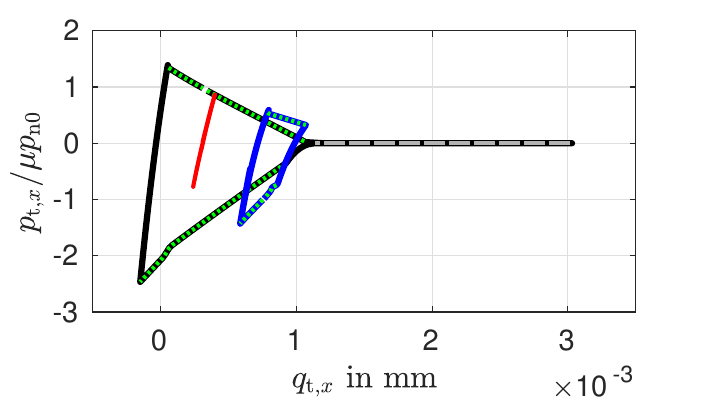}
    \caption{}
    \end{subfigure}
    \hfill
    \begin{subfigure}[b]{.45\textwidth}
    \centering
    \includegraphics[width=\textwidth]{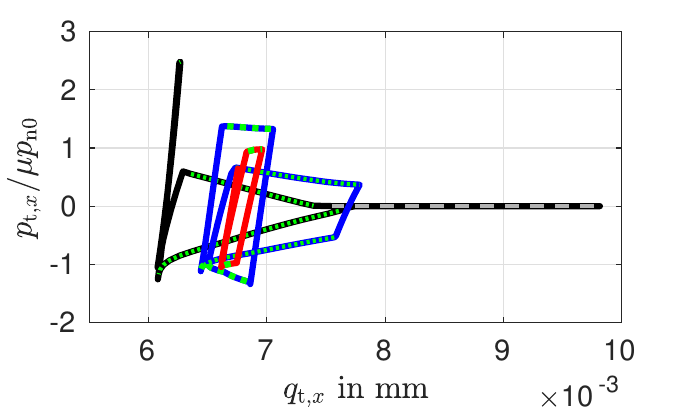}
    \caption{}
    \end{subfigure}
    \caption{
    Representative results of the steady-state contact behavior: Time evolution of tangential relative displacement (top) and frictional hysteresis cycles (bottom) of contact nodes 203, 204 and 205 indicated in \fref{FEmodel}c, at amplitude indicated in \fref{natural_frequencies}g-h for the geometrically linear (left) and the geometrically nonlinear (right) case. Dotted green and gray lines indicate slipping and separation, respectively; otherwise the contact nodes are in stick.
    }
    \label{fig:cstates}
\end{figure}
\\
In \fref{natural_frequencies}a-b, the amplitude-dependent frequency $\omega_1$ and damping ratio $D_1$ are depicted, for the case of tied contact.
As the contact is permanently sticking, no dissipation occurs, leading to $D_1=0$.
The geometrically nonlinear behavior (bending-stretching coupling) leads to a considerable hardening trend, as expected for an initially flat plate.
The frequency error is $<0.5\%$ for all considered reduced models.
This is of the same order of magnitude as in linear case.
\\
Next, geometrically linear behavior is considered, and nonlinear contact behavior is activated.
The results are presented in \fref{natural_frequencies}c-d for the case of low initial pressure.
As bending-stretching coupling is neglected, there is no significant hardening.
In fact, a mild softening trend is observed, which is typical for frictional contact.
Again, the frequency error is $<0.5\%$ for all considered reduced models, consistent with the error already present in the linear regime.
The overall damping level is relatively low.
\\
Finally, geometrically nonlinear and nonlinear contact behavior are considered simultaneously.
The results are presented in \fref{natural_frequencies}e-f for the case of high, and in \fref{natural_frequencies}g-h for low initial pressure.
Compared to the case of tied contact, the hardening trend saturates beyond a certain amplitude.
Here, some of the axial clamping stiffness is lost due to sliding friction.
Apparently, this saturates the membrane stretching and, thus, the bending stiffness.
This is an important effect of the nonlinear contact behavior on the geometrically nonlinear behavior.
Again, the frequency error is $<0.5\%$ for all considered reduced models, consistent with the error already present in the linear regime. Also, the damping ratio is in excellent agreement.
\\
Interestingly, the frictional damping is higher in the geometrically nonlinear case.
This is an important effect of the geometrically nonlinear behavior on the nonlinear contact behavior.
To understand this, representative results of the contact behavior are illustrated in \fref{cstates}, for the geometrically linear case (a and c) and the geometrically nonlinear case (b and d), for low initial pressure.
It is useful to note that the modal damping ratio $D_1$ is determined as the energy dissipated per cycle, divided by $4\pi$ times the maximum potential energy, $(\omega\hat\eta)^2/2$ (\eref{DQSMA}).
Results are shown for equal potential energies; the corresponding displacement amplitudes at the panel center are indicated by green $+$-markers in \fref{natural_frequencies}g-h.
For the selected potential energy, the damping ratio is four times higher in the geometrically nonlinear case ($D_1=0.32\%$ instead of $D_1=0.08\%$).
The results are depicted in terms of the steady-state time evolution of the relative tangential displacement (\fref{cstates}-a and b) and the normalized tangential pressure vs. displacement hysteresis cycles (\fref{cstates}-c and d), for the three contact nodes, 203, 204 and 205, indicated in \fref{FEmodel}c.
At the given potential energy level, the nodes 202 and 201 are sticking and thus not shown in \fref{cstates}.
Thanks to the symmetry in the $y$-direction, the three contact nodes 203, 204 and 205 are representative for the entire contact area, and since the results correspond to the same potential energy, the area enclosed in the frictional hysteresis cycles is representative of the damping ratio.
In general, the results obtained from the finite element and the reduced model are in very good agreement with regard to the contact behavior.
To avoid that \fref{cstates} becomes overcrowded, only results obtained from the reduced-order model are shown.
\\
There are similarities but also important differences between the geometrically linear and nonlinear cases in \fref{cstates}.
In both cases, the contact node at the panel edge separates within the half cycle where the panel bends in positive $z$-direction, $0\leq \tau \leq \pi$.
In the geometrically nonlinear case, much longer sliding distances occur.
Also, the geometrically nonlinear behavior yields two clearly visible stretching-compression cycles per bending cycle.
In particular, one can see the formation of two pronounced local maxima of $g_{\mrm t,x}$, one near $\tau=\pi/2$, and one near $\tau=3\pi/2$, for each of the three depicted contact nodes in \fref{cstates}b.
This also leads to the double hysteresis cycle for contact node 204 in \fref{cstates}d, which is not present in the geometrically linear case (\fref{cstates}c).
Starting from essentially zero tangential pressure in the non-deformed (flat) configuration, bending leads to a stretching load in cycle-average.
This leads to a positive mean value of $g_{\mrm t,x}$.
At the given potential energy, it took as many as 20 load cycles to reach a constant mean position and a steady hysteresis cycle.
\\
In summary, good agreement was achieved already for $ M=1$.
The accuracy is slightly higher for $ M=3$.
Apparently, not more and not less than $ M_\Gamma=3$ interface constraint modes are needed.
In that case, the constraint modes are associated with axial and lateral interface translation, and interface rotation around the bending axis.
A remaining error can, in principle, be due to the truncation of the component modes, or the truncation of the Taylor series expansion of the geometrically nonlinear terms.
However, for the considered vibration regime, this remaining error is regarded as sufficiently small.

\subsection{Response to impulsive loading\label{sec:impulse}}
The proposed model reduction approach should be able to account for the nonlinear interaction of different modes.
To analyze this, the response to an impulsive loading is considered in this subsection.
Starting from the static equilibrium, a single period of a sinusoidal force was applied,
\ea{
{\mm f}_{\mrm{ext}} =
\begin{cases}
-{\mm M}\mm b A \sin\left(2\pi \frac{t}{T}\right) & 0\leq t\leq T \\
\mm 0 & T\leq t
\end{cases}\fp
}
The load pattern mimics an imposed base acceleration in the $z$-direction.
Accordingly, the elements of the vector $\mm b$ are 1 if the corresponding nodal degree of freedom is aligned with the $z$-direction, and 0 if it is orthogonal ($x$- and $y$-direction).
The duration of the sine pulse was set to $T=0.6\cdot2\pi/\omega_{2,\mrm{lin}}$.
From \tref{linfreq}, one can infer that $\omega_{2,\mrm{lin}}/(2\pi)=447.3~\mrm{Hz}$ and $\omega_{1,\mrm{lin}}/(2\pi)=82.5~\mrm{Hz}$.
Thus, the sine pulse is expected to provide significant energy to both the fundamental and the second mode.
The second mode should more precisely be referred to as second odd-ordered bending mode; torsion and even-ordered bending modes are not relevant due to the symmetry of the structure and its loading.
Three acceleration amplitudes which cover the range from almost linear to strongly nonlinear behavior were tested.
The acceleration amplitudes ($A$) are specified in the caption of \fref{disp_dynamic}.
For the simulation, implicit numerical time step integration was employed, using Newmark's constant-average-acceleration scheme.
A constant step size $\Delta t= T/1000$ was used.
The simulation was run until $t=140T$.
\begin{figure}[ht!]
    %
    \begin{subfigure}[b]{.49\textwidth}
    \centering
    \includegraphics[width=\textwidth]{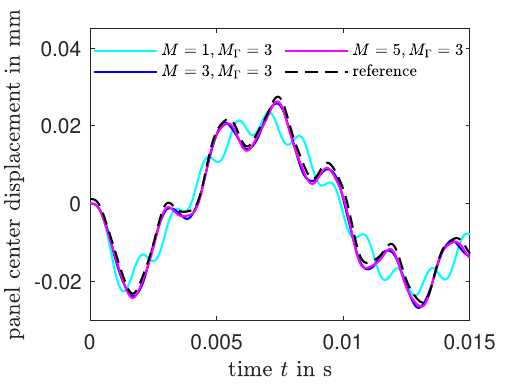}
    \caption{}
    \label{fig:disp_low_conv}
    \end{subfigure}
    \hfill
    \begin{subfigure}[b]{.49\textwidth}
    \centering
    \includegraphics[width=\textwidth]{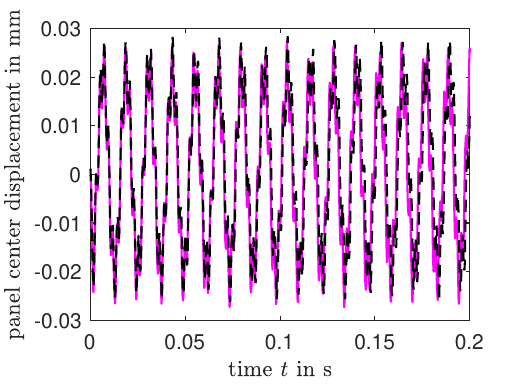}
    \caption{}
    \label{fig:disp_low}
    \end{subfigure}
    %
    \begin{subfigure}[b]{.49\textwidth}
    \centering
    \includegraphics[width=\textwidth]{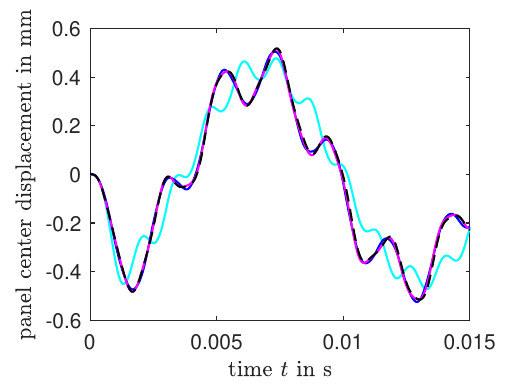}
    \caption{}
    \label{fig:disp_medium_conv}
    \end{subfigure}
    \hfill
    \begin{subfigure}[b]{.49\textwidth}
    \centering
    \includegraphics[width=\textwidth]{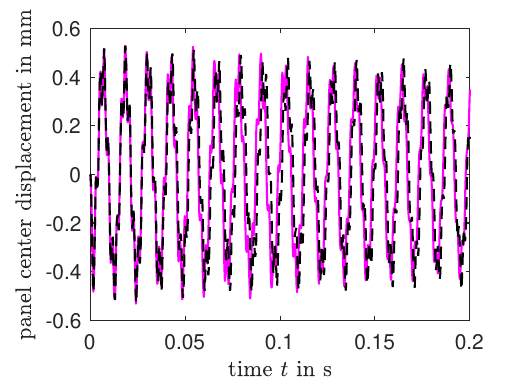}
    \caption{}
    \label{fig:disp_medium}
    \end{subfigure}
    %
    \begin{subfigure}[b]{.49\textwidth}
    \centering
    \includegraphics[width=\textwidth]{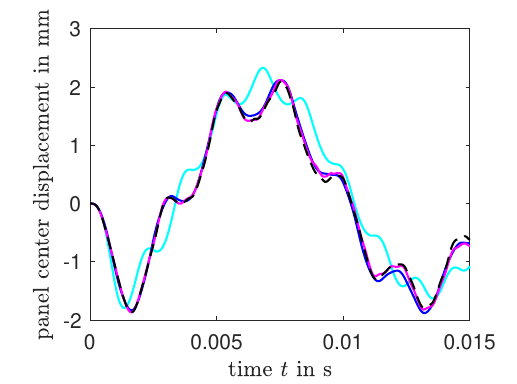}
    \caption{}
    \label{fig:disp_high_conv}
    \end{subfigure}
    \hfill
    \begin{subfigure}[b]{.49\textwidth}
    \centering
    \includegraphics[width=\textwidth]{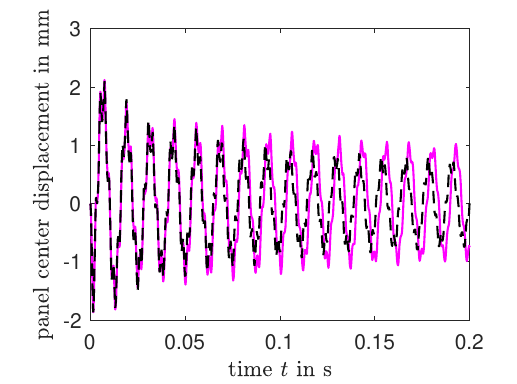}
    \caption{}
    \label{fig:disp_high}
    \end{subfigure}
    \caption{
    Displacement response at the panel center to impulsive loading. The right column shows the complete simulated time span, the left column the initial phase only. The rows correspond to different load levels: (a) and (b) $A=\SI{50}{\meter\per\square\second}$; (c) and (d)  $A=\SI{1,000}{\meter\per\square\second}$; (e) and (f)  $A=\SI{4,000}{\meter\per\square\second}$.
    }
    \label{fig:disp_dynamic}
\end{figure}
\\
The time evolution of the displacement response at the panel center is shown in \fref{disp_dynamic}.
As expected, the participation of more than one modal frequency is clearly visible.
Since no linear damping was specified, the response may only decrease due to frictional dissipation.
For the low load level (top row in \fref{disp_dynamic}), the frictional clamping is mainly sticking, so that the vibrations do not significantly decay.
For the medium and the high load level (middle and bottom row in \fref{disp_dynamic}), apparently, more frictional dissipation occurs and the vibrations decay slowly and more rapidly, respectively.
During the initial $0.015~\mrm{s}$ (left column in \fref{disp_dynamic}), excellent agreement of the reduced model with the reference is observed.
For the low and the medium load level, $M=3$ with $M_{\Gamma}=3$ is sufficient to achieve modal convergence.
For the high load level, increasing the number of normal modes to $M=5$ increases the accuracy visibly.
When considering the complete simulated time span (right column in \fref{disp_dynamic}), some noteworthy discrepancy arises.
In particular, this is the case for the high load level, where the finite element reference decays more quickly.
This cannot be explained with the modal damping results in \fref{natural_frequencies}, where the damping ratio according to the finite element reference is in very good agreement with the proposed method and actually slightly smaller.
We attribute this to numerical damping, which is a well-known problem of dynamic contact simulation using finite element models, see \eg \cite{TRCprediction}.
\begin{table}[ht!]
    \centering
    \caption{Computation wall time. FOM stands for full-order finite element model, ROM for reduced-order model.}
    \begin{tabular}{ccccc}
    \hline
         & FOM & \multicolumn{3}{c}{ROM} \\
         & & $ M_\Gamma=3, M = 1$ & $ M_\Gamma=3, M = 3$ & $ M_\Gamma=3, M = 5$\\
         \hline \hline
       ROM construction & $-$ & $2~\mrm h$ & $10~\mrm h$ & $27~\mrm h$\\
       low-level impulse & $82.46~\mrm h$ & $0.31~\mrm h$ & $0.33~\mrm h$ & $0.35~\mrm h$ \\
       medium-level impulse & $279.98~\mrm h$ & $0.63~\mrm h$ & $0.38~\mrm h$ & $0.60~\mrm h$ \\
       high-level impulse & $285.14~\mrm h$ & $0.61~\mrm h$ & $0.64~\mrm h$ & $0.67~\mrm h$ \\
       \hline
    \end{tabular}
    \label{tab:comp_times}
\end{table}
\\
%
The computational effort required for the finite element analysis is listed in \tref{comp_times} together with that required for the proposed reduction method.
The following soft-/hardware was used:
\ABAQUS 2023, run on 18 cores;
\MATLAB: R2023a, run on up to 18 cores;
Intel Core i9-10980XE @ 3.00 GHz, 36 CPUs;
128 GB RAM.
The computation time for the simulation of the response to impulsive loading is about three orders of magnitude smaller for the reduced model.
The construction of the reduced model generates non-negligible computational overhead (\cf \tref{comp_times}, first row).
In average, a static load case required 4.5 minutes on 18 cores.
It is useful to note that the static load cases can be computed in parallel.
To reduce the overall effort for constructing the reduced model in the future, it could be interesting to consider a reduction basis enriched with modal derivatives in combination with an intrusive procedure for evaluating the polynomial coefficients, instead of the implicit condensation pursued in the present work.
It should be remarked that the overall speedup, accounting for both the simulation and the overhead for constructing the reduced model, depends on how many simulations / dynamic load scenarios are to be analyzed.

\subsection{Replacement of the support region\label{sec:replace}}
\begin{figure}
    \centering
    \begin{subfigure}{0.46\textwidth}
        \includegraphics[width=\textwidth]{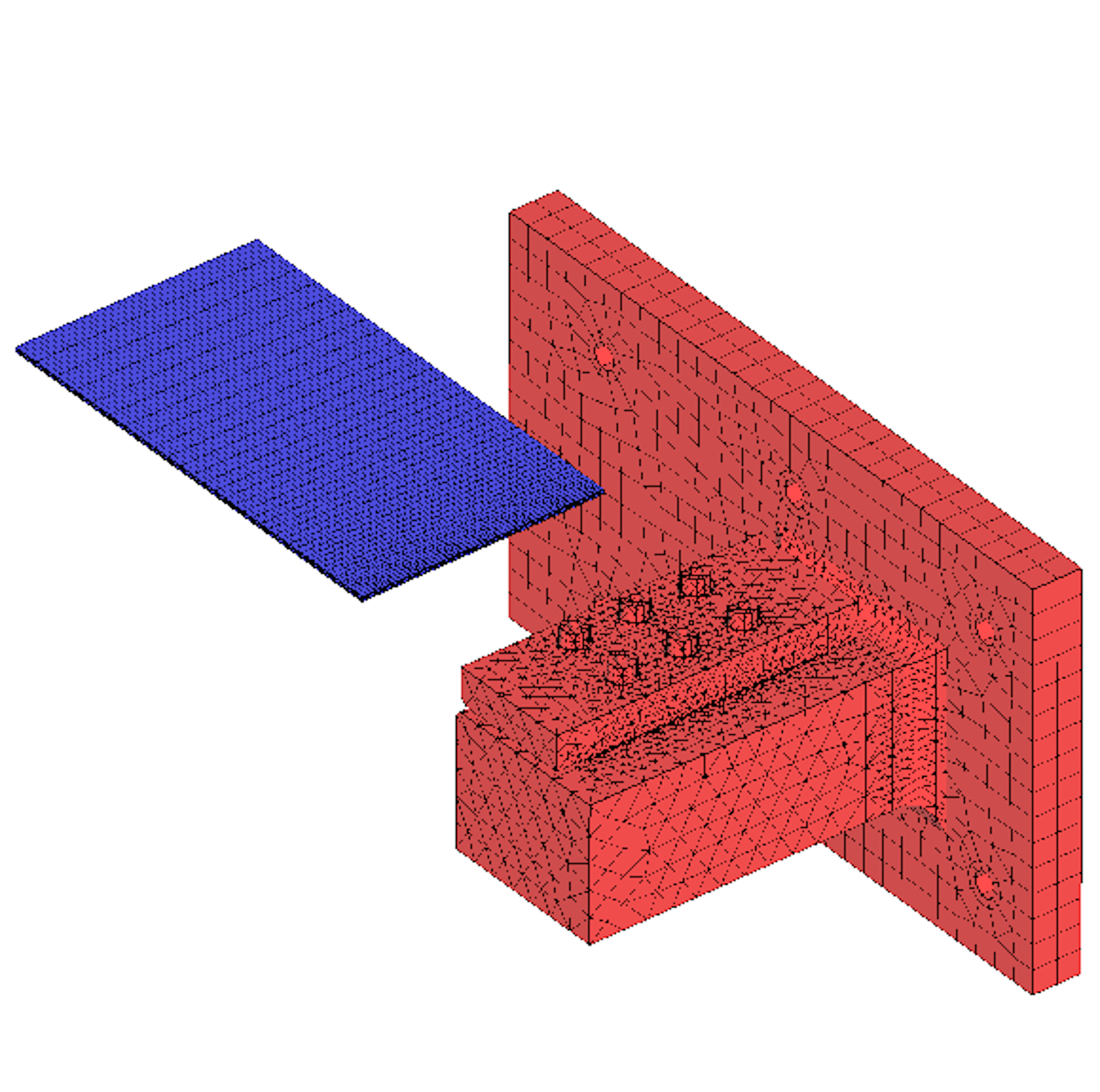}
        \caption{}
        \label{fig:substruct_bolted}
    \end{subfigure}
    \hspace{0.5cm}
    \begin{subfigure}{0.46\textwidth}
        \includegraphics[width=\textwidth]{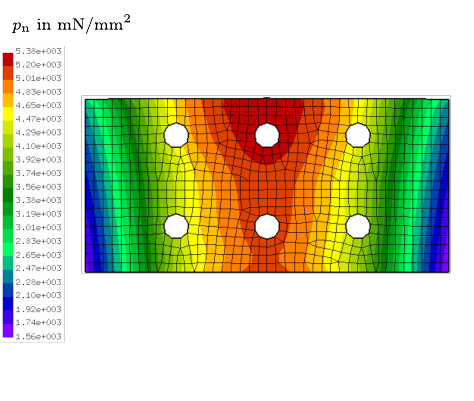}
        \caption{}
        \label{fig:preload_bolted}
    \end{subfigure}
    \caption{Modified benchmark problem with a refined support model containing bolts and bore holes: (a) division of finite element model into thin-walled region (blue) and support region (red); (b) initial contact pressure distribution in the bolted contact interface.}
    \label{fig:model_bolted}
\end{figure}
To illustrate the modular character of the proposed sub-structuring approach, a refined support model is considered in this subsection.
Bolts and bore holes are now included, and also a blade, and a thick base plate are included.
The panel is sandwiched between blade and pillar, and the pillar forms one monolithic piece with the base plate.
With this, the geometry is even closer to the benchmark system in \cite{TRCprediction,Muller.2022c}. 
As the same panel is still considered, the \emph{reduced model of the thin-walled region did not have to be re-computed} but was adopted from the previous analysis.
Only the reduced model of the new support region had to be computed.
To model the bolt tightening, a $2~\mrm{kN}$ force was applied to the end face of each stud, and a force of the same magnitude but opposite direction was applied to the bottom of the corresponding blind hole.
This modeling approach is state of the art.
It should be remarked that the specified tightening load is relatively low; ca. $9~\mrm{kN}$ seem more appropriate for the given bolt type \cite{TRCprediction}.
The intent behind the relatively low bolt load was to provoke a stronger mutual interaction between geometrically nonlinear and nonlinear contact behavior, in order to test the limitations of the proposed approach.
Contact was modeled between panel and pillar with 838 node-to-surface contact elements.
In this modified problem, the contact model parameters were specified as $k_\mrm{n} = 10^6~\mrm{N}/\mrm{mm}^3$, $g_{\mrm{sl}} = 10~\mum$, and $\mu=0.3$.
\fig[htb]{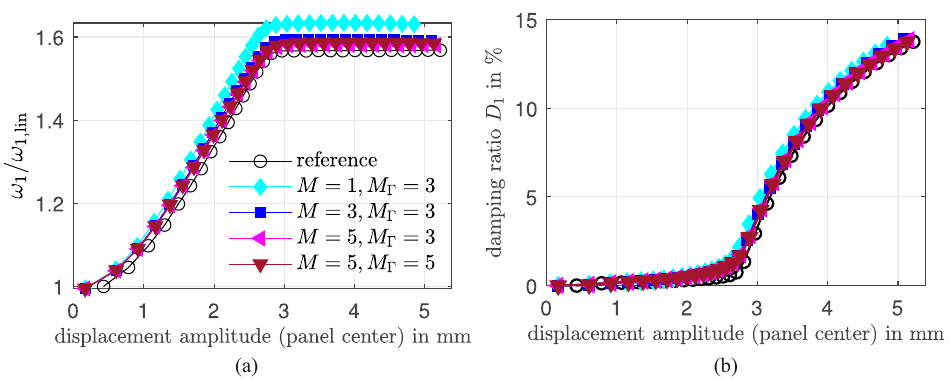}{Amplitude-dependent modal frequency (a) and damping ratio (b) for the modified benchmark problem.}{1.0}
\\
The resulting initial pressure distribution is shown in \fref{model_bolted}b.
Note that the pressure distribution is highly non-uniform in contrast to that in \fref{FEmodel}c.
The amplitude-dependent frequency and damping ratio of the fundamental bending mode are shown in \fref{QSMA_bolted}.
The frequency is normalized by the linear one of about $90~\mrm{Hz}$.
The qualitative evolution is similar: initial hardening induced by bending-stretching coupling, followed by a stiffness saturation and a damping increase when friction becomes important at higher amplitudes.
Again, the agreement of the proposed approach with the finite element reference is deemed very good.
For a higher number of interface and normal modes, the results of the proposed approach stabilize rapidly.
Upon close inspection, one may notice a small offset in frequency, similar to the previous variant of the benchmark problem.
Also, the reduced model transitions more smoothly and slightly too early into the gross slip regime than the finite element reference.

\section{Conclusions\label{sec:conclusions}}
A model reduction approach based on sub-structuring was proposed for thin-walled jointed structures.
The key idea is to divide the system into thin-walled and support regions, so that geometrically nonlinear and nonlinear contact behavior are separated, and available methods for reducing the component models can be employed.
Implicit condensation was used for modeling the geometrically nonlinear behavior within the thin-walled region.
Here, a systematic, engineering-oriented solution was proposed to the open and delicate problem of finding appropriate load scales.
The only user-defined parameters are a target displacement and a stress limit.
The validity of the proposed model reduction approach was numerically demonstrated, with regard to amplitude-dependent modal properties and the response to impulsive loading.
Interesting insight into the mutual interaction of geometrically nonlinear and contact behavior was gained:
On the one hand, an increasing amount of sliding in the clamping limits the extent of bending-stretching coupling.
On the other hand, the geometrically nonlinear bending-stretching deformation has an appreciable effect on the extent of frictional dissipation.
Consequently, those two sources of nonlinear behavior should be considered together in order to properly describe the dynamics of thin-walled jointed structures.
An important benefit of the proposed approach is the modular setup: the parameters of individual components can be varied, and even entire reduced component models can be replaced with minimal effort.
\\
The key limitation of the proposed model reduction approach is inherited from implicit condensation, which neglects nonlinear inertia effects within the reduced basis.
Further, since the number of required static load cases grows rapidly, implicit condensation becomes prohibitive when many (\eg many more than 10) component modes are needed.
This is expected to become relevant, in particular, when the substructure interface section deforms more severely than in the considered examples.
It would then be interesting to use an intrusive enriched-basis technique for reduced modeling of the thin-walled regions as alternative to implicit condensation.

\section*{Acknowledgements}
M. Krack is grateful for the funding received by the Deutsche Forschungsgemeinschaft (DFG, German Research Foundation) [Project 450056469].


\appendix

\section{Construction of interface modes according to Carassale and Maurici\label{asec:OPS}}
In this appendix, the construction of $\mm \Gamma$ in \erefs{supportHCB} and \erefo{constraintModesG} according to \cite{Carassale.2017} is described.
In general, the same basis vectors are used for each of the three translational degrees of freedom:
\ea{
\mm \Gamma = \lbrace v_{\ell,k} \rbrace \otimes \mm I_{3\times 3}\fk\label{eq:threeDOF}
}
where $\otimes$ denotes the Kronecker product, $\mm I_{3\times 3}$ is the three-dimensional identity matrix, and $v_{\ell,k}$ is the $k$-th base function, evaluated at node $\ell$.
Here, it is assumed that the initial finite element model is three-dimensional.
If it is two-dimensional, we have two translational degrees of freedom, and $\mm I_{2\times 2}$ has to be used in \eref{threeDOF} instead.
\\
The $v_{\ell,k}$ are obtained by a three-step procedure:
\begin{enumerate}
    \item Define elementary polynomial terms (degree $P$): $x^ay^b \quad a,b\geq 0,\, a+b\leq P$, where $x$, $y$ are Cartesian coordinates, with origin at the center of the interface area.
    \item Use the Gram-Schmidt process to successively obtain orthogonal polynomials $k=1,2,\ldots$. To this end, use the finite element mesh to evaluate the required inner products (\eg using Gauss point quadrature).
    \item Evaluate the polynomial terms at the node location $(x_\ell,y_\ell)$ to obtain $v_{\ell,k}$.
\end{enumerate}
It is assumed that each node within the interface area can be parameterized uniquely by the coordinates $x$, $y$.
If the interface is not flat, a curvilinear coordinate system could be useful.

\section{Estimation of polynomial coefficients within implicit condensation via regression\label{asec:regression}}
The polynomial coefficients $\lbrace\beta_{2,i}^{jk}\rbrace$, $\lbrace\beta_{3,i}^{jkl}\rbrace$ in \eref{NLgeomTAYLOR} are obtained by regression to the set of static load cases described in \ssref{IC}.
More specifically, one requires that the static response of the reduced model to the considered load cases, agrees well with that obtained from the finite element model.
The load cases are generally expressed as $\mm K\mm T\mm w$ (\cf \eref{staticProblemIC}).
The static response of the reduced model is governed by $\tilde{\mm K}\tilde{\mm q} + \tilde{\mm f}_{\mrm{geom}}(\tilde{\mm q}) = \tilde{\mm f}_{\mrm{ext}}$ (\cf \eref{tROM}).
Herein, $\tilde{\mm f}_{\mrm{ext}}$ is the projection of the static load $\mm K\mm T\mm w$, $\tilde{\mm f}_{\mrm{ext}}=\mm T^{\mrm T}\mm K\mm T\mm w=\tilde{\mm K}\mm w$.
The static response of the finite element model is obtained as the solution of \eref{staticProblemIC}.
The corresponding response in reduced coordinates is estimated as $\tilde{\mm q}=\mm T^+\mm q$ where $\mm T^+$ is the Moore-Penrose inverse of $\mm T$.
This leads to the requirement that
\ea{
\tilde{\mm f}_{\mrm{geom}}\left(\mm T^+\mm q\right) \stackrel{!}{=} \tilde{\mm K}\left(\mm w - \mm T^+\mm q\right)\fp\label{eq:regressionPerLoadCase}
}
Herein, $\mm T^+\mm q$ is obtained for the given $\mm w$ for each load case.
$\tilde{\mm K}$ is available from component mode synthesis.
$\tilde{\mm f}_{\mrm{geom}}\left(\mm T^+\mm q\right)$ is linear in the coefficients $\beta_{2,i}^{jk}$, $\beta_{3,i}^{jkl}$ (\cf \eref{NLgeomTAYLOR}).
The equations for all load cases are combined.
Yet, one obtains independent problems for each row (index $i$ in \eref{NLgeomTAYLOR}).
In each row, we have $R^3/6 + R^2 + 5R/6$ coefficients.
On the other hand, the number of load cases is $4R^3/3 - 2R^2 + 8R/3$, where $R=M+M_\Gamma$ is the number of component modes.
Here, positive and negative load scales are considered, both in the single-mode load cases and in the multi-mode load cases.
With this, the problem is over-determined with respect to the sought polynomial coefficients; the least-squares estimate is used.

\section{Assessment of the proposed load scaling for implicit condensation\label{asec:ICassessment}}
\fig[htb]{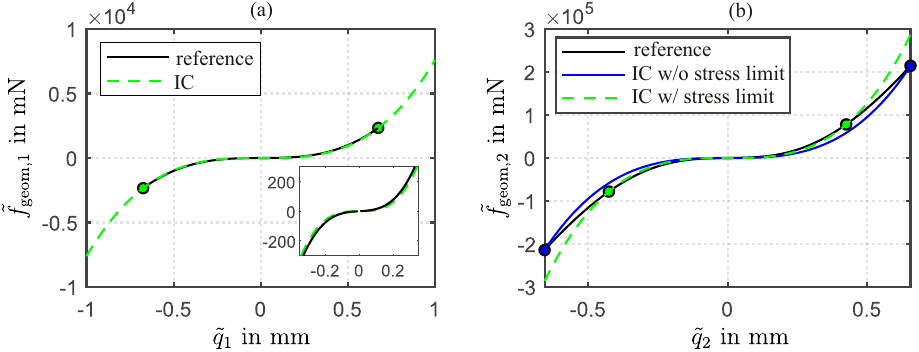}{
Implicit condensation for representative single-mode load cases where no buckling occurs: Nonlinear force-displacement relation for (a) first, (b) second fixed-interface bending mode depicted in \fref{fims}(a) and (b), respectively. IC stands for implicit condensation. The filled circular markers indicate the reference points for regression.
}{1.0}
In this appendix, representative results for the implicit condensation are shown.
The aim is to assess the accuracy of the reduced model for the geometrically nonlinear behavior obtained with implicit condensation, and, in particular, the importance of the proposed load scaling.
Recall that the present state of the art is to specify a displacement (reached according to linear theory) only, but no stress limit nor buckling limit.
\\
In \fref{fit_FIM}, results of two single-mode load cases are shown for which \emph{no buckling occurs}.
The respective modes are the first and second fixed-interface normal mode.
For the first mode, the stress limit was not reached before the target displacement.
The fit obtained with implicit condensation is in excellent agreement with the reference in the depicted range.
If one zooms out, considering larger displacements, it becomes obvious that a cubic-degree polynomial becomes insufficient and higher-order terms would be needed to improve accuracy.
For the second mode, the stress limit was reached before the target displacement.
Within the stress limits, the proposed load scaling leads to a better agreement with the finite element reference.
%
\fig[htb]{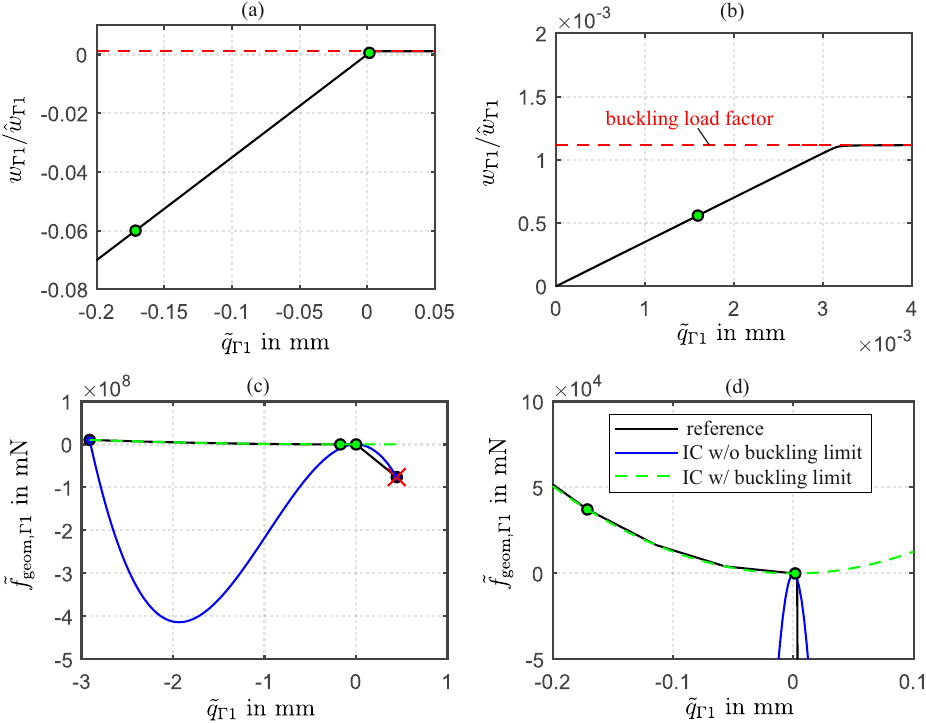}{
Implicit condensation for a representative single-mode load case where buckling occurs: (a,b) Load scale factor vs. modal displacement; (c,d) nonlinear force-displacement relation for the first panel interface mode depicted in \fref{infc}(a). IC stands for implicit condensation. The filled circular markers indicate the reference points for regression. Red cross indicates point where finite element analysis aborted.
}{1.0}
\\
In \fref{fit_IMbuckling}, results of the single-mode load case corresponding to the first panel interface mode (\fref{infc}a) are shown for which \emph{buckling occurs}.
This mode is associated with axial interface movement and hence induces membrane stretching/compression.
First, we wanted to verify the buckling load factor, $\gamma_{\mrm{crit},\Gamma 1}$, obtained from a linear analysis.
To this end, a nonlinear buckling analysis was carried out.
To ensure that the corresponding stable post-buckled equilibrium was reached, a small geometric imperfection was introduced in the shape of the lowest buckling mode, which was also obtained from the aforementioned linear analysis.
The imperfection was scaled so that the maximum absolute shape deviation was $1~\mum$.
The results in \fref{fit_IMbuckling}a-b demonstrate that the buckling load factor obtained with linear analysis is accurate.
Note that the compression part is associated with positive interface displacement ($\tilde q_{\Gamma 1}>0$).
Next, the nonlinear force-displacement relation and its regression were analyzed.
The results are shown in \fref{fit_IMbuckling}c-d.
In the tensile part, the reference curve resembles a parabola.
Beyond the buckling point, the curve deviates severely from this parabola.
The relation looks quasi-linear in the post-buckled regime, which is linked to the fact that a different equilibrium (post-buckled) state has been reached, which is associated with a different (and lower) effective stiffness, leading to a nonlinear force-displacement relation with negative slope.
Clearly, this intricate relation cannot be properly captured by any cubic-degree polynomial.
With the proposed load scaling, implicit condensation accurately captures the pre-buckling regime.
Without the proposed buckling limit, no reliable regression can be expected.

\section{Quasi-static modal analysis\label{asec:QSMA}}
The method for estimating the amplitude-dependent modal frequency and damping ratio, employed in \ssref{NMA}, is described in this appendix.
The implementation follows closely that presented in \cite{festjens2013,Allen2016QuasistaticMA}.
The theory is described for the application to the finite element model (\eref{FOM}).
The application to the reduced model in \eref{systemROM} is analogous.
\\
Suppose that the mode order $k$ is to be analyzed.
First, the equations of the assembled system are linearized around the static equilibrium, and the mass-normalized mode shape $\mm\phi_{k,\mrm{lin}}$ is determined,
\ea{
\left({\mm K} + \left.\partial {\mm h}/\partial {\mm q}\right|_{\mm 0} - \omega_{k,\mrm{lin}}^2{\mm M}\right)\mm\phi_{k,\mrm{lin}} &=& \mm 0\fk \\
\mm\phi_{k,\mrm{lin}}^{\mrm T}{\mm M}\mm\phi_{k,\mrm{lin}} &=& 1\fp
}
Then, a load is applied,
\ea{
{\mm f}_{\mrm{ext}} &=& {\mm M}\mm\phi_{k,\mrm{lin}}\alpha\fp
}
The load is varied quasi-statically.
Consequently, inertia forces are neglected in \eref{FOM}, ${\mm M}\ddot{{\mm q}}\approx\mm 0$.
The modal coordinate $\eta$, frequency $\omega$, and damping ratio $D$ are estimated as
\ea{
\eta &=& \mm\phi_{1,\mrm{lin}}^{\mrm T}{\mm M}{\mm q}\fk \label{eq:etaQSMA}\\
\hat \eta &=& \frac{\max\eta - \min\eta}2\fk \label{eq:etahatQSMA}\\
\omega &=& \sqrt{\frac{\hat\alpha}{\hat\eta}}\fk \label{eq:omQSMA}\\
D &=& \frac{E_{\mrm{diss}}}{2\pi\left(\omega\hat\eta\right)^2}\fp \label{eq:DQSMA}
}
Herein, $\max\eta$ and $\min\eta$ are the maximum and minimum of $\eta$, respectively, over the steady hysteresis cycle obtained for load amplitude $\hat \alpha$.
The denominator in \eref{DQSMA} corresponds to $4\pi$ times the maximum potential energy over the cycle, $(\omega\hat\eta)^2/2$.
\\
Two different variants of the described method have been implemented and used in the present work.
For the results presented in \ssref{NMA}, the steady hysteresis cycle was determined for many different load levels ($\hat\alpha$) and used for evaluating \erefs{etaQSMA}-\erefo{DQSMA}, whereas in \ssref{replace}, an approximation based on the initial loading curve was used.
In the former variant, a load is applied as $\alpha(\tau) = \hat{\alpha}\sin\tau$ for 20 cycles of the pseudo-time variable $\tau$, $0\leq \tau/(2\pi)\leq 20$.
In most cases, a steady state was reached at that point; \ie, the last cycle is closed.
The dissipated energy can then be easily determined from the last cycle as $E_{\mrm{diss}} = \oint {\mm h}^{\mrm T}\dd {\mm q}$.
In the latter variant, the two branches of the initial loading curve are determined, both starting from the equilibrium, one going to some maximum $\hat\alpha$ and the other to $-\hat\alpha$.
One then simply has $\max\eta = \eta(\hat\alpha)$ and $\min\eta=\eta(-\hat\alpha)$.
The energy dissipated per cycle was approximated as in \cite{festjens2013}.
The variant based on the steady hysteresis cycle is associated with much more computation effort, but it is expected to yield more meaningful results in the case of asymmetric hysteresis cycles and severe normal load variation.

\end{document}